\documentclass[a4paper,11pt]{article}
\pdfoutput=1

\usepackage{jcappub} 
\usepackage[T1]{fontenc} 
\usepackage{lineno}
\usepackage{tabularx}
\usepackage{booktabs}
\usepackage{cleveref}
\usepackage{mathtools}
\usepackage{graphicx}
\usepackage{subcaption}
\usepackage{orcidlink}

\title{\boldmath Building a Better Beta: Nucleation and Timescales in Cosmological Phase Transitions}

\author[1]{William Searle\note{Corresponding author.},\orcidlink{0009-0000-2392-9689}}
\author{and Csaba Bal\'azs,\orcidlink{0000-0001-7154-1726}}
\affiliation{School of Physics and Astronomy, Monash University,\\
Melbourne 3800 Victoria, Australia}

\emailAdd{william.searle@monash.edu.au}
\emailAdd{csaba.balazs@monash.edu.au}

\arxivnumber{2608.30115}


\abstract{
First-order phase transitions in the early universe can generate a stochastic background of gravitational waves, offering a unique probe of high-energy physics. 
In this work, we investigate aspects of bubble nucleation and transition timescales, which play a central role in shaping the resulting gravitational wave spectrum. 
Many common approaches characterise the transition rate via a Taylor expansion of the false vacuum decay rate.
We argue that a more fundamental description is instead given by the distribution of bubble lifetimes, and define a new timescale, $\beta_\nu$, as the first moment of this distribution. 
We show that $\beta_\nu$ reproduces the behaviour of previous timescale definitions in the appropriate limits, while avoiding their pathologies, and offers a more natural description of the ensemble of nucleated bubbles. 
We then quantify the impact of this improved timescale on the predicted gravitational wave spectrum, finding that it shifts the peak amplitude by up to an order of magnitude relative to previous definitions.
As next-generation gravitational wave detectors come online, robust theoretical predictions will be essential; we hope this work represents a step in that direction.
}

\newcommand{\code}{\texttt}
\newcommand{\textsub}[2]{#1_\text{#2}}

\newcommand{\lrbrace}[1]{\left ( #1 \right )}

\newcommand{\ms}{\overline{\text{MS}}}

\newcommand{\veff}{V_{\text{eff}}}
\newcommand{\vext}{\mathcal{V}_\text{ext}}
\newcommand{\phim}{\phi_{\text{m}}}

\begin{document}
\maketitle
\flushbottom

\section{Introduction}
\label{sec:intro} 

The early universe was home to a variety of rich and interesting phenomena. Many of these processes, such as cosmological inflation or the mergers of primordial black holes, produce a stochastic gravitational wave background (SGWB). This includes cosmological phase transitions (PT), which can leave behind a variety of observable relics; from topological defects \cite{Kibble:1976sj}, primordial black holes \cite{Hawking:1982ga, Bond:1984tga}, to a SGWB itself \cite{Witten:1984rs, Hogan:1896yaj}. Unlike conventional electromagnetic radiation, gravitational waves (GW) are decoupled from the Standard Model (SM) at the Planck scale, and as such, the universe is virtually transparent to these GW signals, meaning they can still carry information about these physical processes today. Several future GW observatories, such as LISA \cite{LISA:2017pwj}, Taiji \cite{Hu:2017mde}, TianQin \cite{TianQin:2015yph}, DECIGO \cite{Musha:2017usi}, and BBO \cite{Corbin:2005ny}, will be sensitive to these signals. This motivates a better understanding of the physical processes that source these GWs, as well as the development of the theoretical tools and computational pipelines required to predict their spectra. Much of our understanding of the universe before Big Bang Nucleosynthesis (BBN) remains an extrapolation of physics tested at much lower energies. A direct observation of a PT would push this understanding to energy scales otherwise inaccessible to collider experiments.

A first-order cosmological phase transition (FOPT) can be realised in a variety of beyond-the-Standard-Model (BSM) theories. Grand unified theories require a sequence of PT as the unified symmetry group is spontaneously broken to the SM gauge group. Alternatively, additional scalar fields can result in a scalar sector that predicts a much richer thermal history, with new intermediate phases similarly providing new pathways through which to reach the SM vacuum. Furthermore, a dark sector featuring one or more dark Higgs fields can also predict a strong FOPT \cite{Schwaller:2015tja}, providing an attractive probe of elusive dark matter physics. As the early universe cools, a FOPT proceeds through the nucleation of bubbles of the new, more stable phase \cite{Kamionkowski:1993fg}. The released energy from the metastable phase then drives the expansion of these bubbles, which can source GWs through several mechanisms. The bubble wall interface itself can generate GWs due to the anisotropic stress of the scalar field configuration \cite{Kosowsky:1991ua, Kosowsky:1992vn, Jinno:2016vai, Jinno:2017fby}. Interactions between the bubble wall and the surrounding plasma generate sound waves that propagate throughout the medium after the walls have collided, producing GWs from this bulk motion \cite{Hindmarsh:2013xza, Hindmarsh:2015qta, Hindmarsh:2017gnf, Hindmarsh:2016lnk, Hindmarsh:2019phv}. Finally, the formation of non-linear features in this bulk motion produces a distinct turbulence spectrum of GWs \cite{Caprini:2006jb, Caprini:2009yp}. Of these contributions, numerical simulations indicate that the acoustic component is typically the dominant source of GWs for electroweak-scale transitions.

The acoustic GW power spectrum depends on a chain of complex interdependent calculations, starting from a derivation of the effective potential, to a calculation of the false vacuum decay rate and thermal history, to thermodynamics, hydrodynamics, and so on. Many approaches to this calculation, such as the sound shell model (SSM) \cite{Hindmarsh:2016lnk, Hindmarsh:2019phv, RoperPol:2023dzg}, distil this chain down to a small set of thermal parameters, including the transition strength, $\alpha$, the bubble wall velocity, $v_w$, as well as characteristic time and length scales, $\beta$ and $L$. One does this at the cost of losing information from the effective potential and introducing significant degeneracy in the GW inverse problem\footnote{Consider, for example, the case where the number of Lagrangian parameters is larger than the number of thermal parameters.}. For example, $\alpha$ allows one to replace the full equation of state (EoS) with an approximate description using the bag model \cite{Steinhardt:1981ct, Kosowsky:1992rz, Kamionkowski:1993fg, Megevand:2009ut, Leitao:2010yw, Konstandin:2010dm, Espinosa:2010hh, Megevand:2012rt, Leitao:2015fmj, Leitao:2015ola}. However, this simplified EoS can result in qualitatively different outcomes for the fluid profiles when compared to the calculation using the full EoS. These differences can be as severe as predicting different solutions for the hydrodynamical mode \cite{Giese:2020rtr, Giese:2020znk, Wang:2020nzm, Tian:2024ysd, Linton:2026fpj}.

Similar to the EoS, another important input for the acoustic spectrum, and indeed for all contributions to the GW spectrum, is the description of bubble nucleation. This is contained in the false vacuum decay rate, $\Gamma(t)$. This determines how bubble nucleation proceeds and, for the acoustic contribution, sets the structure of the initial fluid velocity field. The decay rate is typically approximated as being either exponential or Gaussian \cite{Hindmarsh:2019phv}. Exponential transitions nucleate rapidly, producing an ensemble of bubbles whose maximum radii are set by the initial population of nucleated bubbles, with further bubbles able to nucleate rapidly in any available false vacuum, populating the smaller radii. Conversely, Gaussian transitions produce an ensemble of bubbles whose radii are dominated by the population that nucleates around the peak decay rate. In both cases, the full decay rate is replaced with an approximation derived from a single parameter, $\beta$, which is obtained from an instantaneous evaluation of the decay rate. This approximation can often lead to arbitrary and incorrect predictions for the GW spectrum \cite{Athron:2023rfq}.

A clear analogy can be drawn between these two cases. Qualitatively, using $\alpha$ or the bag EoS replaces the full description of the physics with a single instantaneous point estimate of the transition strength. This fails to preserve the information contained in the effective potential and leads to a less consistent description of the system. This can be remedied by replacing the bag EoS with the full EoS. By analogy, we wish to replace $\beta$ with a full description of the bubble nucleation derived from the effective potential. The most natural candidate is the false vacuum decay rate itself. However, we argue that instead a more suitable choice is the distribution of bubble lifetimes. Instead of just describing the decay rate, this distribution describes the physical population of nucleated bubbles and is directly used in the calculation of the acoustic spectrum in the SSM. Just as a more self-consistent calculation of the spectrum replaces $\alpha$ with the full equation of state, we aim to replace $\beta$ with the full lifetime distribution. Furthermore, we also introduce a new timescale, $\beta_{\nu}$, which is the average lifetime determined from this distribution. We show this yields results equivalent to those of other candidates for the timescale but suffers from none of the pathologies associated with these alternatives.

The calculations performed in this work use a combined numerical pipeline consisting of \code{PhaseTracer2} \cite{Athron:2024xrh}, \code{BubbleDet} \cite{Ekstedt:2023sqc}, and \code{HydroGrav} \cite{Linton:2026fpj}\footnote{A notable omission is \code{WallGo} \cite{Ekstedt:2024fyq}. The wall velocity does not appear in the decay rate and only slightly affects the false vacuum fraction. As such, we use an approximation of $v_w$ in place of the full calculation. The only place \code{WallGo} would reasonably improve the findings of this paper would be in the GW spectrum calculation in \cref{sec:effects_on_gws}. However, in this comparison, we only need to assume a constant $v_w$ across each case.}. \code{PhaseTracer2} performs the phase-tracing of the effective potential, as well as the calculation of the bounce action in the computation of the false vacuum decay rate. In addition, we extend the functionality of the base \code{PhaseTracer2} to perform a fully self-consistent calculation of the false vacuum fraction. \code{BubbleDet} allows for precise calculation of the false vacuum decay rate, including contributions from the functional determinant, whilst \code{HydroGrav} performs the calculation of both fluid profiles and the resulting acoustic power spectrum using the SSM. Combined, these three codes provide a state-of-the-art prediction for the GW spectrum. Notably, this calculation no longer depends on either $\alpha$ or $\beta$, with the former being replaced by the full EoS and the latter by the lifetime distribution.

The remainder of this paper is outlined as follows. In \cref{sec:models} we introduce the two benchmark models used throughout: the real scalar singlet extension of the SM, and a classically conformal $U(1)$ DM model. In \cref{sec:thermal_history}, we detail our treatment of the thermal history of a phase transition, including our calculation of the false vacuum decay rate and approach to the self-consistent false vacuum fraction. In \cref{sec:nucleation_and_timescales}, we introduce both the exponential and Gaussian descriptions of bubble nucleation, then, in \cref{sec:better_beta}, we introduce our general description of bubble nucleation and define the average bubble lifetime. In \cref{sec:effects_on_gws}, we study the effects these improvements have on the observable GW power spectrum, and we give our conclusions in \cref{sec:conclusions}.

\section{Benchmark Models}
\label{sec:models}

The first benchmark model considered in this work is the $\mathbb{Z}_2$ real scalar singlet extension, xSM. This predicts a first-order phase transition driven by a real scalar singlet coupled to the Higgs~\cite{Choi:1993cv, Ham:2004cf}, whose tree-level potential is
\begin{equation}
    V(H, S) = -\mu_h^2 H^\dagger H + \lambda_h (H^\dagger H)^2 - \frac{\mu_s}{2} S^2 + \frac{\lambda_s}{4} S^4 + \frac{\lambda_{hs}}{2} H^\dagger H S^2, \label{eq:Veff-xSM}
\end{equation}
where the Higgs doublet and scalar singlet are
\begin{equation}
    H = \frac{1}{\sqrt{2}} \begin{pmatrix} G^{\pm} \\ \phi_h - h - iG^0 \end{pmatrix}, \quad S = \phi_s + s.
\end{equation}
Here, $\phi_h$ and $\phi_s$ are background fields, whilst $h$ and $s$ are the dynamic degrees of freedom of the Higgs and scalar fields, respectively, and $G^{\pm,0}$ are the Goldstone bosons. After symmetry-breaking, we are left with two singly charged  $G^\pm$ and a neutral CP-odd $G^0$ Goldstone bosons, corresponding to the electroweak $W^\pm$ and $Z$ bosons. In addition, we have the CP-even scalars $h$ and $s$, with masses:
\begin{equation}
    m_h^2 = 2 v_h^2 \lambda_h, \quad m_s^2 = - \mu_s^2 + \frac{1}{2} v_h^2 \lambda_{hs},
    \label{eq:xsm_masses}
\end{equation}
where $h$ is the SM Higgs, and we have assumed the EWSB vacuum $(\phi_h, \phi_s) = (v_h, 0)$. Constraining $\mu_h$ and $\lambda_h$ using the observed Higgs mass and vacuum expectation value (VEV), this model features three free parameters: the physical scalar mass, $m_s$, where we exchange $\mu_S$ for the physical scalar mass using \cref{eq:xsm_masses}, and the quartic couplings, $\lambda_s$ and $\lambda_{hs}$. The selection of points in this parameter space will be outlined below, but we consider regions where the final vacuum is reached by a two-step transition
\begin{equation}
    (\phi_h, \phi_s) \rightarrow (0, v_s) \rightarrow (v_h, 0),
\end{equation}
and we focus on the second step.

The second benchmark model considered is a classically conformal Abelian Higgs model featuring a $U(1)'$ dark sector, appearing in \cite{Balan:2025uke}. The tree level potential for the dark Higgs, $\Phi$, is
\begin{equation}
    V(\Phi) = \lambda (\Phi^* \Phi)^2.
\end{equation}
Included in the Lagrangian is the dark gauge boson, $A_\mu'$, together with two dark fermions $\chi_{1,2}$ that are oppositely charged under $U(1)'$. After spontaneous symmetry breaking, we have:
\begin{equation}
    \Phi = \frac{1}{\sqrt{2}} ( \phi_b + \phi + i \varphi),
\end{equation}
where $\phi_b$ is the background field, and $\phi$ and $\varphi$ are the dynamical degree of freedom and Goldstone boson, respectively. The model is then determined by three free parameters: the $U(1)'$ gauge coupling, $g$, the dark fermion Yukawa coupling, $y$, and the $\phi_b$ VEV, $v$.\footnote{We require the one-loop potential have minima at $0$ and $v$, which imposes a constraint on $\lambda$ as a function of $g$ and $y$.}

For brevity, we relegate details on the construction of the effective potential for both models to \cref{app:effective_potential}, but in general, we utilise the one-loop Coleman-Weinberg potential with finite temperature corrections \cite{Coleman:1973jx, Dolan:1973qd}. Infrared divergences at finite-temperature are regulated using daisy resummation in the Parwani scheme \cite{Parwani:1991gq}.

For each model parameter space, we identify a one-dimensional slice as follows. For the xSM, we fix $m_S = 125$ GeV and $\lambda_s = 1.0$. Then, we uniformly sample 100 $\lambda_{hs}$ values over the range $(1.040, 1.062)$. This fine-tuning is to select parameter points that undergo a slow transition. We further identify a benchmark point $\lambda_{hs} = 1.061$, which we label BP1. This is taken to be towards the upper-most limit of the possible parameter space and demonstrates a slow, Gaussian transition. Conversely, for the conformal $U(1)$ model, we take $v = 100$ MeV and $y = 0.1$. We then sample 100 $g$ values over the range $(0.5, 1.0)$. Due to the nature of classically-conformal models, we require very little fine-tuning to achieve strong supercooling. The benchmark point, $g = 0.72$, hereafter referred to as BP2, is taken to fall roughly in the middle of this range. These parameters are summarised in \cref{tab:model_params}.

\begin{table}[ht]
\centering
\begin{tabularx}{0.7\textwidth}{c|cccc}
\toprule
Model & $m_s$ ($v$) & $\lambda_s$ ($y$) & $\lambda_{hs}$ ($g$) & BP \\
\midrule
xSM & 125 & 1.0 & (1.040, 1.062) & 1.061 \\
Conformal $U(1)$ & 100 & 0.1 & (0.5, 1.0) & 0.72 \\
\bottomrule
\end{tabularx}
\caption{Model parameters considered for the two benchmark models. The labels correspond to the xSM (conformal $U(1)$) parameters. All dimensional quantities are given in units of GeV for the xSM, and MeV for the conformal $U(1)$ model. The BP column refers to the value of $\lambda_{hs}$ $(g)$.}
\label{tab:model_params}
\end{table}

\section{The Dynamics of Nucleation and Thermal History}
\label{sec:thermal_history}

The thermal history of a phase transition can be measured in a few quantities. The rate at which bubbles form is measured by the false vacuum decay rate, $\Gamma(t)$. The progress of the transition can be seen in the evolution of the true vacuum fraction, $f(t)$, or equivalently the false vacuum fraction, $h(t)$. The total energy and momentum densities, $\bar{\rho}(t)$ and $\bar{p}(t)$, and their effects on the expansion of the universe can be seen in the evolution of the EoS parameter, $w(t)$. Alternatively, the expansion itself is tracked by quantities such as the Hubble rate $H(t)$ or scale factor $a(t)$. To consistently study cosmological phase transitions, each of these quantities should be evolved consistently with one another. In the remainder of this section, we introduce the relevant details of these calculations, with a particular focus on the false vacuum decay rate and fraction, as well as the Friedmann-JMAK equations that couple these quantities together. We introduce relevant theory, methods, and perform demonstrations of each calculation using the benchmark models previously introduced.

\subsection{False vacuum decay rate}
\label{sec:false_vacuum_decay_rate}

The physics of bubble nucleation is largely determined by the false vacuum decay rate. For a given cosmological phase transition, this is defined as the tunnelling rate per unit volume per unit time. It takes the following functional form:
\begin{equation}
    \Gamma(t) = A(t) e^{- S_d(t)},
    \label{eq:gamma_def}
\end{equation}
where $A(t)$ and $S_d(t)$ are herein referred to as the prefactor and bounce action, respectively. In general, $\Gamma(t)$ is given by a functional integral over a set of fields $\phi$ with action $S_d(\phi)$, such that:
\begin{equation}
    \Gamma(t) \sim \int \mathcal{D} \phi e^{-S_d[\phi]}.
\end{equation}
The functional form in \cref{eq:gamma_def} is obtained by performing a saddle-point approximation to this integral, such that $S_d(t)$ comes from the classical action evaluated at the saddle point, whilst $A(t)$ represents fluctuations. Accurate evaluation of both $A(t)$ and $S_d(t)$ remains an active problem in the study of FOPT, with several publicly available codes now available for calculating it across a variety of algorithms \cite{Masoumi:2017trx, Wainwright:2011kj, Athron:2019nbd, Sato:2019wpo, Guada:2020xnz, Ekstedt:2023sqc, Basler:2024aaf, Athron:2024xrh, Costa:2025pew}.

The bounce action, $S_d(t)$, is the Euclidean action evaluated for the $O(d)$-symmetric bounce solution, $\phi_b(\rho)$ \cite{Coleman:1977py}:
\begin{equation}
    S_d(t) \equiv S_E[\phi_b] = \frac{(2 \pi)^{d/2}}{\Gamma(d/2)} \int_0^{\infty} d \rho \rho^{d-1} \left [ \frac{1}{2} \left ( \frac{d \phi_b}{d \rho} \right)^2 + V(\phi_b )\right ],
    \label{eq:bounce_action}
\end{equation}
where $\rho^2 = \tau^2 + |\vec{x}|^2$ and $\tau$ is the Euclidean time. The bounce path can be obtained by solving the classical equations of motion:
\begin{equation}
    \frac{d^2 \phi}{d \rho^2} + \frac{d-1}{\rho}\frac{d \phi}{d \rho} = V'(\phi),
\end{equation}
subject to the boundary conditions:
\begin{equation}
    \lim_{\rho \rightarrow \infty} \phi_b(\rho) = \phi_F, \quad \frac{d \phi_b}{d \rho} \Bigg |_{\rho = 0} = 0.
\end{equation}
To calculate the bounce path and then the action, we utilise \code{PhaseTracer2} \cite{Athron:2024xrh}, which uses a version of the path deformation algorithm first developed in \code{CosmoTransitions} \cite{Wainwright:2011kj}. We evaluate the bounce action at finite temperature, in which the full $O(d)$-symmetric bounce solution in Euclidean time is instead replaced by an $O(d-1)$-symmetric solution, where the time coordinate is integrated out. The bounce solution $\phi_b$ then obeys
\begin{equation}
    \nabla^2 \phi_b = V'(\phi_b, T).
\end{equation}
The bounce action in the false vacuum decay rate is then replaced according to $S_d(t) \rightarrow S_{d-1}(T)/T$ \cite{Linde:1977mm}, and the full temperature-dependent effective potential must then be used in the evaluation of \cref{eq:bounce_action}.

The prefactor, $A(t)$, is often assumed to play a minor role in the evolution of $\Gamma(t)$, which is instead thought to be dominated by the behaviour of $S_d(t)$. As such, the following approximation is typically used:
\begin{equation}
    A(T) = T^4 \left ( \frac{S_3(T)}{2 \pi T} \right )^{3/2}.
    \label{eq:approx_prefactor}
\end{equation}
This approximation has been demonstrated to give a poor description of the full decay rate. In truth, the size of $\log(A)$ relative to $S_d$ is akin to the relative contribution of first-order perturbative effects in field theory and can play a decisive role in the decay rate \cite{Ekstedt:2023sqc}. The correct prefactor at finite temperature and to one-loop order is given by the functional determinant \cite{Callan:1977pt, Linde:1980tt}:
\begin{equation}
    A(T) = T \left ( \frac{S_3(\phi_b)}{2 \pi T} \right)^{3/2} \left | \frac{\det'(-\nabla^2 + V''(\phi_b, T))}{\det(-\nabla^2 + V''(\phi_F, T))}\right |^{-1/2},
    \label{eq:correct_prefactor}
\end{equation}
where the $'$ indicates that zero modes have been omitted in the determinant. The functional determinants required to calculate $A(t)$ introduce significant complexity, motivating approximations such as \cref{eq:approx_prefactor}. Fortunately, this issue can be alleviated by using the package \code{BubbleDet}, which provides a first-of-its-kind numerical calculation of the functional determinants required by the correct prefactor. 

Unfortunately, one assumption underlying \code{BubbleDet} is that, despite working for multi-field potentials, it assumes only one background field is coordinate-dependent. This assumption allows for a variety of models to be explored using the package, but in the context of the electroweak phase transition, many BSM theories assume a second scalar field participates in a two-step transition. Of the two models explored in this work, the xSM is incompatible with these assumptions. Given this, we use the approximation in \cref{eq:approx_prefactor} when calculating the false vacuum decay rate for the xSM. For the conformal $U(1)$ model, only one field is participating in the transition of interest, and as such, we use the full expression in \cref{eq:correct_prefactor}.

\begin{figure}[t]
    \centering
    \includegraphics[width=\linewidth]{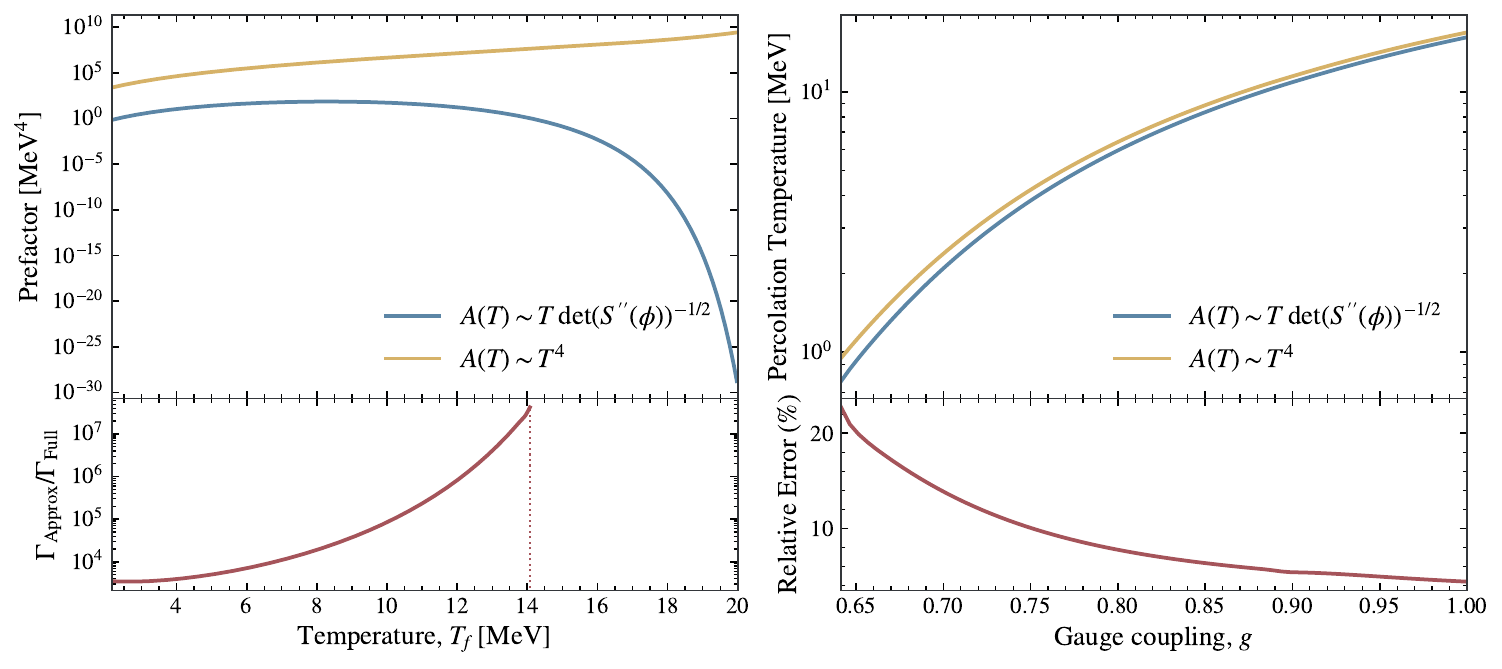}
    \caption{Comparison of the decay rate prefactor $A(T)$ (top left) and the ratio of the full false vacuum decay rate (bottom left) for BP2. The horizontal dotted line indicates the temperature at which $e^{-S_3(T)/T} \approx 0$ to within numerical precision. Also shown is the comparison of the percolation temperature calculation using each prefactor for the conformal $U(1)$ benchmark model (top right), and the relative error (Full - Approx)/Full in the temperature (bottom right).}
    \label{fig:prefactor_comp}
\end{figure}

\Cref{fig:prefactor_comp} compares the results of the false vacuum decay rate calculation when using the full expression for the prefactor, \cref{eq:correct_prefactor}, compared to the approximation \cref{eq:approx_prefactor} for the conformal $U(1)$ model. 
The left column shows the full prefactor as a function of the false vacuum temperature\footnote{See \cref{app:friedmann-jmak_derivation} for an explanation of true and false vacuum temperatures $T_t$ and $T_f$.} for the two approaches. The top plot shows the two $A(T)$ curves, whilst the bottom plot shows the ratio $\Gamma_\text{Approx}/\Gamma_\text{Full}$, calculated for BP2. The horizontal cutoff indicates the region where the increasing bounce action exponentially suppresses $\Gamma \sim e^{-S_3(T)/T} \approx 0$ to within numerical precision. For lower temperatures, the prefactor calculations differ by several orders of magnitude. As we approach the critical temperature, the difference in the prefactor becomes large, approaching $\mathcal{O}(10^{40})$. However, in this region the increasing bounce action suppresses the overall false vacuum decay rate, leading to no practical difference.

In the right column, we propagate the changes in $\Gamma(T)$ to the full calculation of the percolation temperature, defined in \cref{eq:perc_and_comp_def}, for the full conformal $U(1)$ parameter space. The top plot shows each temperature, whilst the bottom gives the relative error (Full - Approx)/Full. The changes in the false vacuum decay rate lead to a percentage-level relative error in the percolation temperature when compared to the approximate calculation, which can become as large as 20\% in extreme cases. Whilst this difference may seem minor, particularly when compared to other theoretical uncertainties \cite{Croon:2020cgk, Athron:2022jyi, Lewicki:2024xan}, the large supercooling present in this classically conformal model makes it extremely susceptible to these small changes. In particular, the transition strength, $\alpha$, can change by orders of magnitude. This comparison is quantified in \cref{tab:alpha_comp}.
\begin{table}[t]
\centering
\begin{tabularx}{0.7\textwidth}{c|cccc}
\toprule
$A(T)$ [MeV$^4$] & $T_n$ [MeV] & $T_p$ [MeV] & $T_f$ [MeV] & $\alpha(T_p)$ \\
\midrule
$T^4$ & 1.110 & 0.951 & 0.855 & 4632 \\
$T \det(S''(\phi))$ & 0.916 & 0.774 & 0.638 & 18260 \\
\bottomrule
\end{tabularx}
\caption{Comparison of the thermal parameters $T_\star$ and $\alpha$ for the parameter point with $g = 0.641$ in the conformal $U(1)$ model. Due to the nature of the classically conformal model, the \% level change in the percolation temperature, $T_p$, manifests in an order of magnitude change in the transition strength, $\alpha$.}
\label{tab:alpha_comp}
\end{table}

\subsection{False vacuum fraction}
\label{sec:false_vacuum_frac}

The false vacuum fraction, $h(t)$, is the fraction of the universe in the false vacuum during a cosmological phase transition. It acts as a measure of the progress of a phase transition. In particular, we can define two temperature milestones directly from $h(T)$. These are the percolation, $T_p$, and completion, $T_f$, temperatures:
\begin{equation}
    h(T_p) = \delta_p, \quad h(T_f) = \epsilon_f,
    \label{eq:perc_and_comp_def}
\end{equation}
where we take $\delta_p = 0.71$ and $\epsilon_f = 0.01$. Other than the critical temperature, we also define the nucleation temperature in terms of the nucleation rate, $N(t)$, as $N(T_n) = 1$\footnote{Throughout, we use time and temperature interchangeably, where they are related through the solutions to the Friedmann-JMAK equations.}.

In a static universe, the false vacuum fraction is determined completely as a function of the false vacuum decay rate, together with the volume of nucleated bubbles, which itself is a function of the bubble wall velocity, $v_w$, and nucleation time. The situation is more complicated in an expanding universe. In this case, the volume of a bubble nucleated at some time $t$ depends not only on $v_w$, but also on the behaviour of the scale factor $a(t)$ between the nucleation time and the observation time. As such, the false vacuum fraction is determined by both the physics of bubble nucleation and the equations governing the homogeneous universe. In \cref{app:friedmann-jmak_derivation}, we derive the system of equations governing this coupled evolution, herein referred to as the Friedmann-JMAK equations\footnote{Whilst \crefrange{eq:FJMAK_1}{eq:FJMAK_3} follow from the Friedmann equation, strictly speaking \cref{eq:FJMAK_4} is not the JMAK equation: $h(t) = \exp( - \vext(t))$ (see also the Avrami equation). However, \cref{eq:FJMAK_4} follows directly from this relationship, as is derived in \cref{app:friedmann-jmak_derivation}. As such, the system is given this name.}:
\begin{align}
    H(t)^2 & = \frac{8 \pi G}{3} \left [ f(t) \rho_t(t) + h(t) \rho_f(t) \right ], \label{eq:FJMAK_1} \\ 
    \dot{\rho}_f(t) & = - 3 H(t) \left [ \rho_f(t) + p_f(t) \right ], \label{eq:FJMAK_2} \\
    \dot{\rho}_t(t) & = - 3 H(t) \left [ \rho_t(t) + p_t(t) \right ] + \frac{\dot{f}(t)}{f(t)} \left [ \rho_f(t) - \rho_t(t) \right ], \label{eq:FJMAK_3} \\
    \dot{I}_n(t) & = n I_{n-1}(t) + (n-3)H(t) I_n(t); \quad 1 < n \leq 3, \label{eq:FJMAK_4} \\
    \dot{I}_0(t) & = \Gamma(t) - 3 H(t) I_0(t), \label{eq:FJMAK_5}
\end{align}
where $h(t)$ is related to $I_3(t)$ by \cref{eq:false_vac_with_I3}, and an explanation of each symbol is given in the opening paragraph of this section. The full thermal history of a given phase transition is fully determined by the solutions of \cref{eq:FJMAK_1} through to \cref{eq:FJMAK_5}.

\begin{figure}[t]
    \centering
    \includegraphics[width=\linewidth]{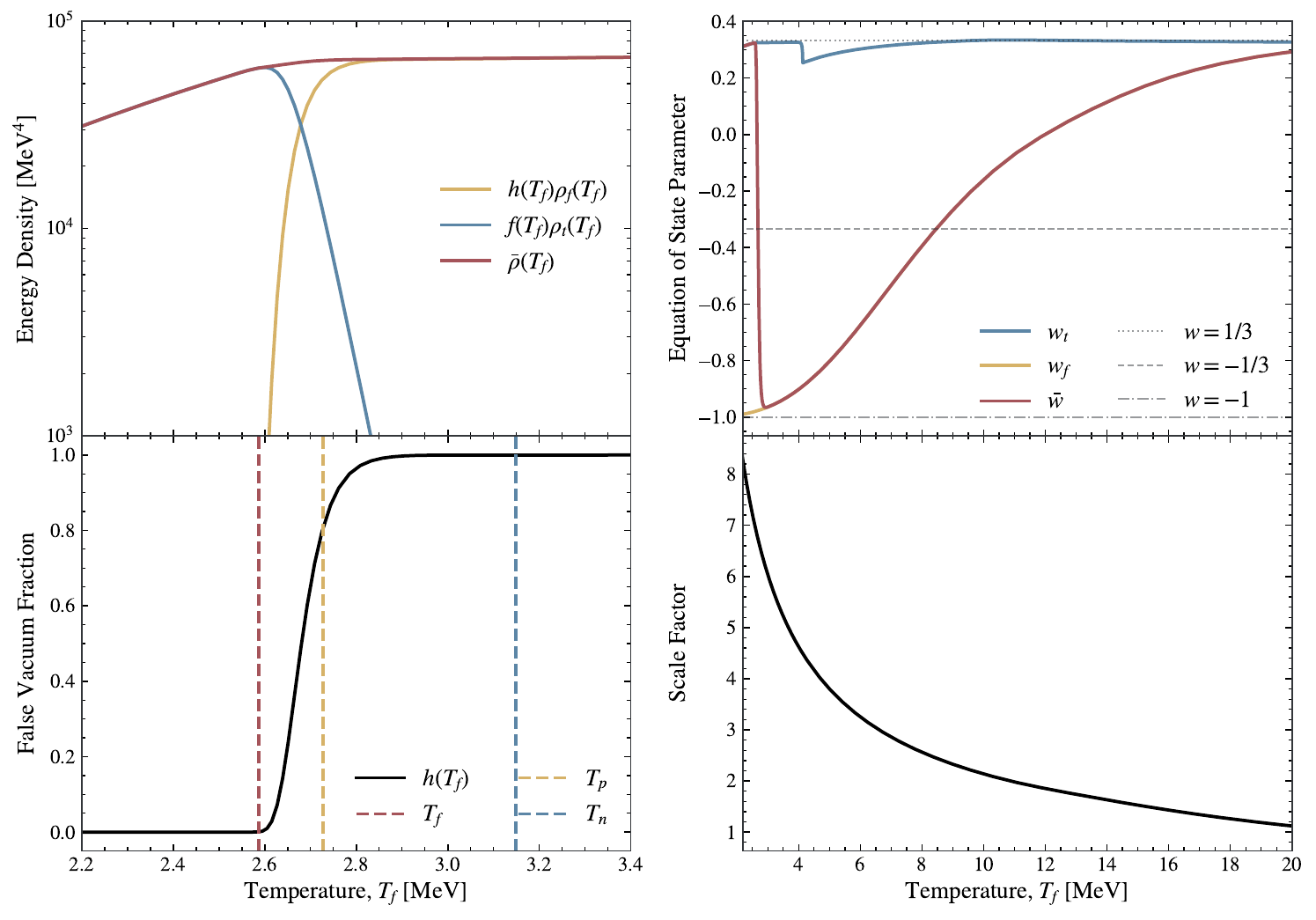}
    \caption{The solution of the Friedmann-JMAK equations for BP2 in the conformal $U(1)$ benchmark model. The top-left cell tracks the evolution of the volume-averaged energy densities, whilst the top-right shows the equation of state, where the particular values for radiation (dotted) and cosmological constant (dash-dotted) are shown, as well as the transition to $\ddot{a} > 0$ (dashed). The bottom-left traces the thermal history, giving the false vacuum fraction and reference temperatures $T_n$, $T_p$, and $T_f$. Finally, the scale factor is shown in the bottom-right. All quantities are given in terms of the false vacuum temperature, $T_f$.}
    \label{fig:thermal_history_u1}
\end{figure}

\begin{figure}[t]
    \centering
    \includegraphics[width=\linewidth]{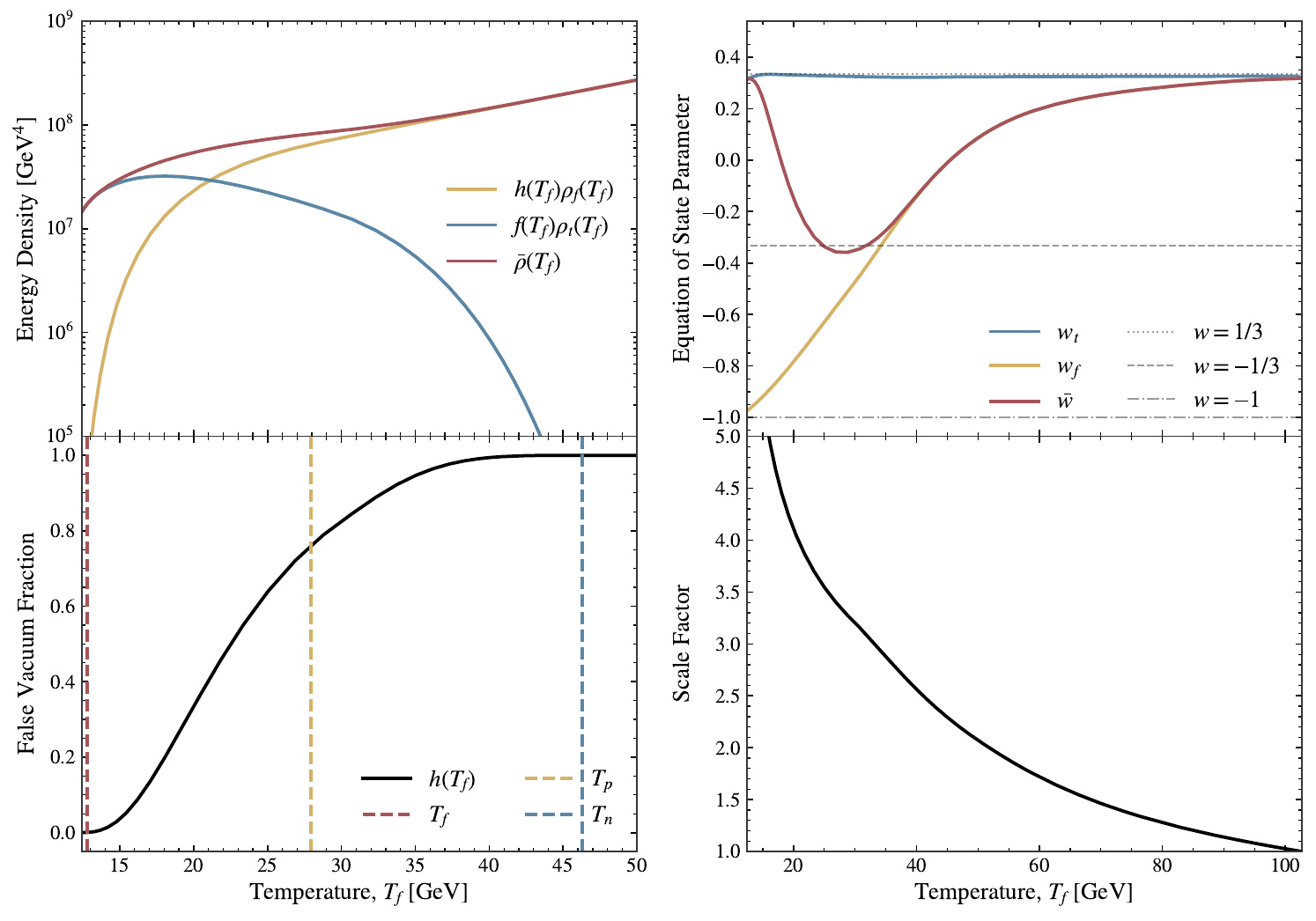}
    \caption{The same plot as \cref{fig:thermal_history_u1}, shown instead for BP1 of the xSM benchmark model.}
    \label{fig:thermal_history_xsm}
\end{figure}

We have implemented this calculation in a modified version of \code{PhaseTracer2}. \Cref{fig:thermal_history_u1} shows the solution of these equations for BP2, while \cref{fig:thermal_history_xsm} shows the solution for BP1. The top-left plot tracks the evolution of the average energy density, indicating the contributions from the volume-averaged true and false vacuums, inspired by a similar plot in \cite{Matuszak:2026xsz}. The bottom-left shows the evolution of the false vacuum fraction, indicating the three reference temperatures $T_n$, $T_p$, and $T_f$. The top-right plot shows the EoS parameter, $w = p/\rho$, again identifying the average and each phase separately. Lastly, the bottom-right plot shows the scale factor. All quantities are shown as a function of the false vacuum temperature, $T_f$. The left column features a tighter $x$-axis, as these quantities typically evolve over shorter scales.

The conformal $U(1)$ model demonstrates large supercooling, $(T_c - T_p)/T_c \rightarrow 1$, and as such enters a period of accelerated expansion with $\bar{w} < -1/3$ throughout a considerable part of the temperature domain. However, the transition still percolates rapidly after nucleating, which can be seen in the sharp transition from the true to false vacuum (best visualised in the plot of $w$), or in the closeness of $T_p$ and $T_n$. This behaviour is contrasted by the xSM benchmark point. This transition exhibits moderate supercooling, $(T_c - T_p)/T_c \sim 0.7$, however, it proceeds much slower than the $U(1)$ case. This can be seen in the slow evolution from false vacuum to true vacuum, or the large separation between the three reference temperatures\footnote{In particular, $(T_p - T_f)/T_c \sim 0.14$ for the xSM, but only $\sim 0.008$ for the $U(1)$ case.}. Of the quantities entering the Friedmann-JMAK system, the false vacuum decay rate, $\Gamma(t)$, is the primary quantity determining this slow nucleation, warranting a more thorough investigation of the decay rate and bubble nucleation.

\section{Nucleation and Timescales}
\label{sec:nucleation_and_timescales}

The nature of bubble nucleation and the evolution of the false vacuum decay rate, $\Gamma(t)$, can be best understood by expanding the bounce action, $S(t)$, about some reference time $t_0$:
\begin{equation}
    S(t) = S(t_0) - \beta_1 (t - t_0) + \frac{1}{2} \beta_2^2 (t - t_0)^2 + \dots.
\end{equation}
The linear term, $S(t_0)$, contributes to the decay rate at $t_0$ and can be absorbed in $\Gamma_0$. The linear and quadratic coefficients are defined:
\begin{equation}
    \beta_1 = - \frac{d S(t)}{dt} \bigg |_{t = t_0}, \quad \beta_2 = \sqrt\frac{d^2 S(t)}{dt^2} \bigg |_{t = t_0}.
    \label{eq:beta_defs}
\end{equation}
The reference time, $t_0$, is conventionally chosen such that $t_0 = t_p$. Both $\beta_1$ and $\beta_2$ contribute differently to the dynamics of bubble nucleation, as explained below.

Firstly, we consider the case in which the transition occurs via thermal tunnelling. In this case, as time increases and the universe cools, the action decreases. If the potential barrier separating the two phases is a product of purely thermal effects, then the action continues to decrease until a minimum temperature, $\textsub{T}{min}$, where the barrier dissipates, is reached. At this point, the action is zero, and the decay rate diverges. Physically, any part of the universe remaining in the false vacuum will undergo a smooth crossover. In this case, the expansion above is well approximated by the linear term. This yields an exponential false vacuum decay rate:
\begin{equation}
    \Gamma(t) = \Gamma_0 e^{\beta_1 (t - t_0)}.
    \label{eq:gamma_exp_in_time}
\end{equation}
This scenario is referred to as exponential nucleation. 

A common approximation for exponential nucleation is to assume that it occurs during radiation domination. In this case, the time-temperature relationship simplifies, and we obtain the common form for the dimensionless ratio $\beta/H$:
\begin{equation}
    \frac{\beta}{H} = - T \frac{d}{dT} \lrbrace{\frac{S_3(T)}{T}}.
    \label{eq:beta_approx}
\end{equation}
However, an exponential transition will not necessarily occur during radiation domination, as was seen in the conformal $U(1)$ model. The transition in \cref{fig:thermal_history_u1} is exponentially nucleated (a claim that will be justified shortly), and yet the transition occurs during a period of vacuum domination for which $\bar{w} < - 1/3$. Under these conditions, the time-temperature relationship above breaks down and the generic \cref{eq:beta_defs} must be used. To maintain generality, we always use the definitions in \cref{eq:beta_defs}, linearising the curve $\Gamma(t)$ obtained by solving the Friedmann-JMAK equations.

We can instead consider the case where the barrier separating the two phases persists at zero temperature. As time increases, the action will decrease until a minimum value is reached at some time $t_m = t(T_m)$. Beyond this, the action begins to increase again, and a local minimum occurs. In this case, the curvature term $\beta_2$ becomes relevant. In particular, if the transition percolates with $t_p \geq t_m$, then $\beta_1 \leq 0$ becomes non-physical, necessitating going beyond linear order. As $\beta_1$ is non-physical, we include only the $\beta_2$ term in the expansion, yielding a Gaussian false vacuum decay rate\footnote{We have written this as an expansion at the point $t_m$, for which $\beta_1$ is exactly zero. Provided $t_p \sim t_m$, we can instead expand around $t_p$ and neglect the $\beta_1$ contribution.}:
\begin{equation}
    \Gamma(t) = \Gamma_m e^{- \frac{1}{2} \beta_2^2 (t - t_m)^2}.
    \label{eq:gamma_gauss_in_time}
\end{equation}
This scenario is correspondingly referred to as Gaussian nucleation. These transitions are characterised by a short period of bubble nucleation, which is then rapidly suppressed by the increasing action. 

Clearly, the choice between $\beta_1$ and $\beta_2$ depends on both quantitative and qualitative features of the bounce action. In each case, the physical role played by $\beta$ is different. The role of $\beta$ depends on how the bubble nucleation proceeds in the transition. For an exponential transition, $\beta_1$ is best understood as a nucleation rate, determining $\Gamma'(t)$. For the Gaussian transition, $\beta_2$ is instead better understood as a width, determining how long the period of bubble nucleation lasts. Furthermore, determining which $\beta$ should be used is itself non-trivial. A minimum in the bounce action implies a peak value for the decay rate, but doesn't necessarily imply that nucleation is Gaussian. If $\Gamma_m/H^4 \gg 1$, then bubble nucleation can reach an appreciable rate long before the minimum in $S(t)$ is ever reached. The nucleation rate is then well approximated by an exponential. As such, Gaussian nucleation typically requires suppression in $\Gamma_m/H^4$, such that bubble nucleation only occurs around the peak of the Gaussian.

\section{Lifetime Distribution and a Better Beta}
\label{sec:better_beta}

A useful way around the intricacies of $\beta_1$ and $\beta_2$ is to define $\beta$ in terms of some other scale in the system. To that end, the following definition is often used:
\begin{equation}
    \beta_\text{eff} = (8 \pi)^{1/3} \frac{v_w}{R_s},
    \label{eq:rs_exp}
\end{equation}
where $R_s$ is the mean bubble separation. This relationship can be derived by assuming static space, $a(t) \approx \text{const}$, and that $\Gamma \sim e^{\beta t}$. Under these assumptions, the integral for the mean bubble number density, $n(t)$, is given by
\begin{equation}
    n(t) = \int_{t_0}^{t_n} d t' \Gamma(t') h(t') \left( \frac{a(t')}{a(t)} \right )^3,
    \label{eq:mean_bubble_density}
\end{equation}
can be evaluated exactly as a function of $v_w$ and $\beta$. Upon assuming $R_s = n^{-1/3}$, the above formula results. The value for $\beta_\text{eff}$ is then expected to agree well with $\beta_1$ provided the transition occurs fast enough for expansion and supercooling to play no relevant role.

Whilst it agrees exactly in the case of fast, exponential nucleation, $\beta_\text{eff}$ can be used generically. To justify this, we consider the case of homogeneous bubble nucleation with mean bubble density $n$. In 3d, the average nearest neighbour distance between nucleation points is given by \cite{Chandrasekhar:1943ws}:
\begin{equation}
    \langle R \rangle = \Gamma \left (\frac{4}{3} \right ) \left (\frac{4 \pi n}{3} \right)^{-\frac{1}{3}} = \Gamma \left (\frac{4}{3} \right ) \left (\frac{3}{4 \pi} \right)^{\frac{1}{3}} R_s.
\end{equation}
We then obtain a timescale, $\beta_\text{eff}^{-1}$, by assuming $\beta_\text{eff}^{-1} = \langle R \rangle / v_w$, which is the time taken for the bubble wall to traverse the average nearest-neighbour distance between nucleation points. This agrees with the previous result up to an $\mathcal{O}(1)$ prefactor. The interpretation of $\beta_\text{eff}$ in this approach is that of a lifetime, where it is assumed that a bubble disappears once its nucleation point is crossed by another bubble.

This discussion motivates the introduction of a more formal description of the lifetime of nucleated bubbles. We achieve this by introducing the distribution over bubble lifetimes, $\nu(t_L)$. This is defined such that the probability that a bubble disappears with a lifetime inside the small interval $[t_L, t_L + dt_L]$ is given by $\nu(t_L) dt_L$. In \cref{app:lifetime_distribution}, we derive the following expression for $\nu(t_L)$:
\begin{equation}
    \nu(t_L) = - \frac{1}{\mathcal{N}_\infty} \int_{t_0}^\infty d t_n \Gamma(t_n) a(t_n)^3\frac{dh}{dt'} \Bigg |_{t' = t_n + t_L},
    \label{eq:lifetime_dist_full}
\end{equation}
where we recall $\Gamma$ is the false vacuum decay rate, $h$ is the false vacuum fraction, $a$ is the scale factor, and $\mathcal{N}_\infty$ is a normalisation factor. Also in \cref{app:lifetime_distribution}, we give the analytical forms for exponential and simultaneous nucleation\footnote{Simultaneous nucleation assumes $\Gamma(t) = \Gamma_0 \delta(t - t_0)$, and can be obtained as a limiting case of Gaussian nucleation.}, respectively:
\begin{equation}
    \nu_\text{exp}(\tilde{t}) = e^{-\tilde{t}}, \quad \nu_\text{sim}(\tilde{t}) = \frac{1}{2} \tilde{t}^2 e^{-\frac{1}{6}\tilde{t}^3},
    \label{eq:nu_approx}
\end{equation}
where the dimensionless time is given by $\tilde{t} = \beta t$.

\begin{figure}[t]
    \centering
    \begin{subfigure}[b]{0.48\textwidth}
        \centering
        \includegraphics[width=\textwidth]{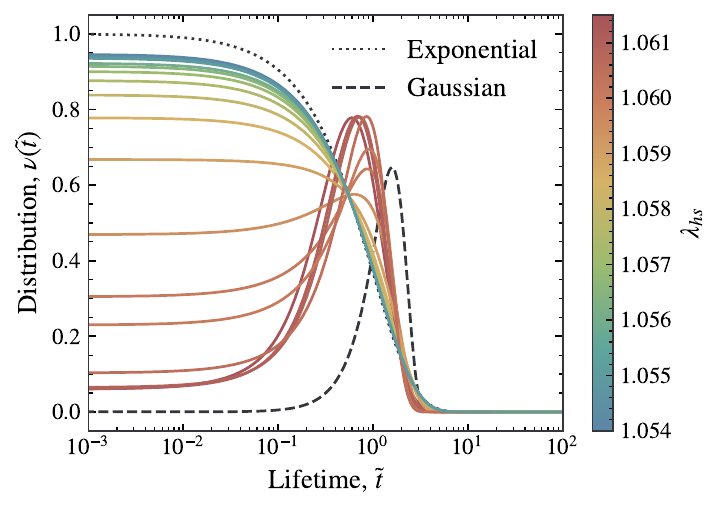}
    \end{subfigure}
    \hfill
    \begin{subfigure}[b]{0.48\textwidth}
        \centering
        \includegraphics[width=\textwidth]{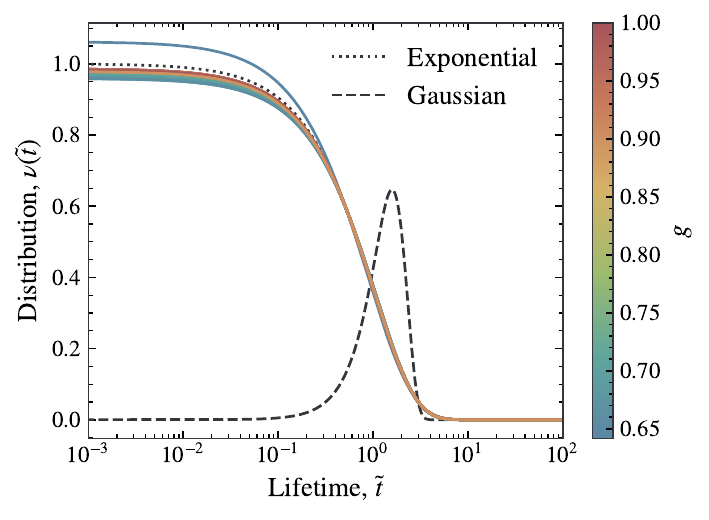}
    \end{subfigure}
    \caption{The bubble lifetime distribution, $\nu(t_L)$, evaluated for the xSM (left) and conformal $U(1)$ (right) benchmark models. Also shown are the exponential (dotted) and Gaussian (dashed) approximations to the distribution. Each curve is shown as a function of the dimensionless parameter $\tilde{t} = \beta t$.}
    \label{fig:dists}
\end{figure}

The expressions for $\Gamma(t)$, $a(t)$, and $dh/dt$ are all given in the solutions to \cref{eq:FJMAK_1} through \cref{eq:FJMAK_5}. As such, despite its intimidating form, we can readily perform the integral in \cref{eq:lifetime_dist_full}. This has once again been implemented in a modified version of \code{PhaseTracer2}. In \cref{fig:dists}, we show the dimensionless lifetime distributions for the xSM and $U(1)$ model. Also on these plots, the analytical forms for both exponential and Gaussian nucleation are shown. This justifies our earlier claim that the $U(1)$ model undergoes exponential nucleation, despite being strongly supercooled, with the full distribution aligning strongly with the exponential approximation. 

A much richer behaviour emerges in the full xSM calculation, where nucleation slowly transitions from exponential to Gaussian across the parameter space. This transition creates intermediate regions where neither extreme adequately captures the full lifetime distribution.

At $\lambda_{hs} \sim 1.059$, a peak in $\nu(\tilde{t})$ first forms, whilst the distribution still has the characteristic tail of an exponential transition. This value of $\lambda_{hs}$ corresponds to the point at which $t_p \sim t_m$ and $\beta_1 \sim \beta_2$. For $\lambda_{hs} > 1.059$, the transition begins to percolate after the peak decay rate, $t_p > t_m$. The peak in the distribution begins to become more prominent, whilst the tail becomes shorter. Physically, this tail forms when $\Gamma_m/H^4 \gg 1$, as in this regime, bubbles will still nucleate rapidly during $t > t_m$, albeit at a reduced rate. Hence, a peak will still be present due to the favourable nucleation conditions at $t \sim t_m$, but outside of these bubbles, smaller and short-lived bubbles can still nucleate rapidly in the remaining region of false vacuum. As such, the tail only disappears completely once $\Gamma_m/H^4 \sim 1$. In this case, the decay rate is only appreciable for $t \sim t_m$, leading to a short burst of bubbles nucleating around this point and a prominent peak in the lifetime distribution. 

Clearly, this hybrid regime contains physics not captured by either the Gaussian or exponential limits. In addition, a full description of bubble nucleation cannot be inferred from either the false vacuum decay rate or the false vacuum fraction alone. Both aspects motivate treating the lifetime distribution as a fundamental quantity describing nucleation.

As expected, the distribution over bubble lifetimes depends on the false vacuum decay rate, $\Gamma$, but it also depends on the rate of change of the false vacuum, $dh/dt$. This derivative is only non-zero whilst the false vacuum is being converted to true vacuum, i.e., during the dynamical part of the transition. In other words, the derivative term $dh/dt$ acts as an integration kernel that ensures only nucleation times $t_n$ for which $t_n + t_L$ falls during the dynamical epoch of the transition will contribute to the distribution. It is for this reason that a minimum in $S(t)$, or equivalently a peak in $\Gamma(t)$, does not necessarily imply Gaussian nucleation. If $\Gamma_m/H^4$ is large, $dh/dt$ will be non-zero well before this peak is reached, and over the relevant integration domain in \cref{eq:lifetime_dist_full} the bounce action will be well approximated by the linear term, leading to exponential-like nucleation.

For these reasons, we introduce a new timescale derived directly from $\nu(t_L)$, which is the (inverse) average bubble lifetime, $\beta_{\nu}$, given by:
\begin{equation}
    \beta_{\nu}^{-1} = \langle t_L \rangle = \int_0^{\infty} dt_L t_L \nu(t_L).
    \label{eq:beta_L_definition}
\end{equation}
One immediate advantage of $\beta_{\nu}$, when compared to $\beta_1$, $\beta_2$, and $\beta_\text{eff}$, is that it is independent of $T_\star$. As such, all uncertainty in $\beta$ arising from systematic uncertainty in the reference temperature does not propagate into $\beta_{\nu}$. The value of $\beta_{\nu}$ is obtained by considering the full thermal history of bubble nucleation, which is contained in the lifetime distribution. Furthermore, utilising our consistent evaluation of $h(t)$ derived from the Friedmann-JMAK system means that all the effects of expansion, reheating, etc., on bubble nucleation are already incorporated into $\beta_{\nu}$. None of the aforementioned ambiguity and intricacy in the calculation of $\beta_1$ and $\beta_2$ enters into $\beta_{\nu}$. The above expression uses the full $\Gamma(t)$ curve, without any approximations or expansions based on Taylor-expanding the bounce action. Finally, there is no need for an \textit{a posteriori} determination of the nucleation type, and hence correct $\beta$ prescription, because such a classification of the nucleation type is already encoded in the lifetime distribution and hence $\beta_{\nu}$.

\begin{figure}[t]
    \centering
    \includegraphics[width=\linewidth]{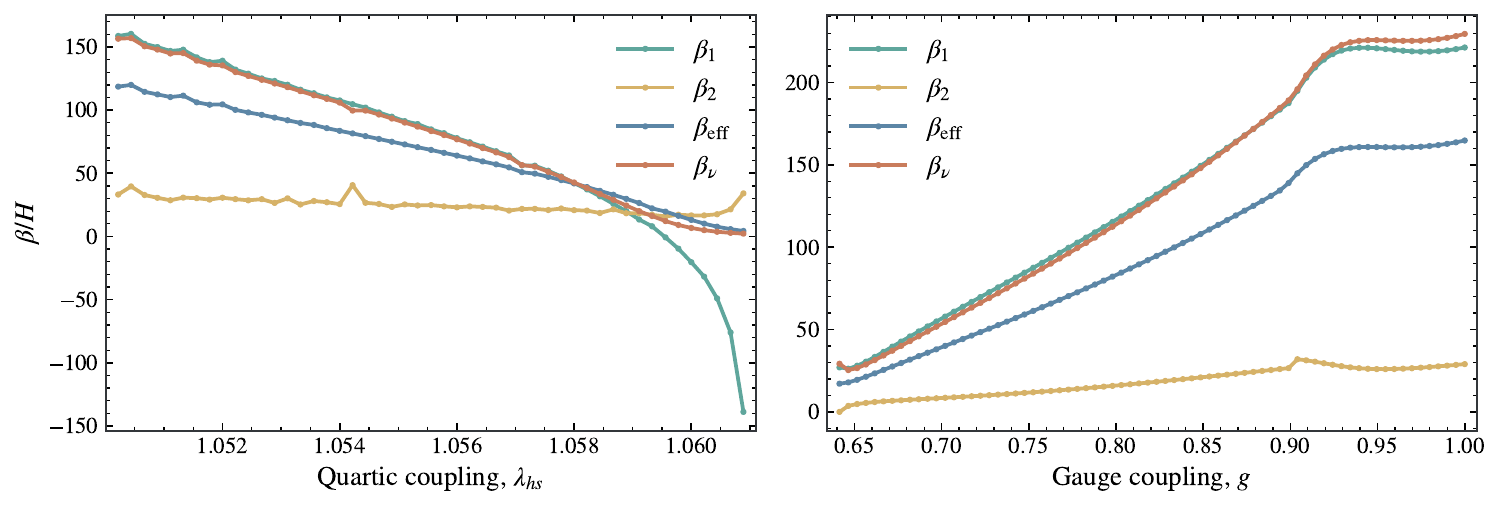}
    \caption{Comparison of the different methods for calculating $\beta$ for the xSM (left) and conformal $U(1)$ (right) benchmark models. Shown are the two timescales derived from Taylor-expanding the action, $\beta_1$ and $\beta_2$, together with the effective timescale $\beta_\text{eff}$, derived from the average bubble separation. Shown in blue is the new timescale, $\beta_\nu$, corresponding to the average bubble lifetime.}
    \label{fig:betas}
\end{figure}

The advantages associated with $\beta_{\nu}$ are for naught if it cannot reasonably replicate the known physics already encoded in $\beta_1$ and $\beta_2$. As such, we note that using the expressions in \cref{eq:nu_approx}, $\beta_{\nu}$ can be evaluated as a function of both $\beta_1$ and $\beta_2$ by inserting the approximate forms in \cref{eq:nu_approx} into the definition \cref{eq:beta_L_definition}. Doing so, we obtain:
\begin{equation}
    \beta_{\nu} = \begin{cases}
        \beta_1, & \text{ exponential nucleation,}\\
        \left [ 6^{1/3} \Gamma \left ( \frac{4}{3} \right) \right ]^{-1} \beta_2 \approx \beta_2,  & \text{ Gaussian nucleation.}
    \end{cases}
    \label{eq:beta_comp_to_old}
\end{equation}
To within $\mathcal{O}(1)$, $\beta_{\nu}$ will replicate the physics of either timescale without requiring an additional determination of the nucleation type. It comes for free in this consistent prescription. To showcase these effects, we calculate each of the four timescales, $\{ \beta_1, \beta_2, \beta_\text{eff}, \beta_{\nu} \}$, for both the xSM and $U(1)$ models in \cref{fig:betas}.

In \cref{fig:betas}, we confirm many of the previous claims made about $\beta_{\nu}$. For the strongly exponential transitions observed in the $U(1)$ model, $\beta_{\nu} \approx \beta_1$, as expected from the analytical result. Alternatively, we see in the xSM that for most of the parameter space, $\beta_{\nu}$ tracks with $\beta_1$, as expected from the exponential nature of these transitions. However, as $\lambda_{hs} \approx 1.059$, we enter the regime of hybrid and Gaussian nucleation. At this point, $\beta_{\nu}$ begins to track with $\beta_\text{eff}$, which, from our earlier discussion, is expected from the interpretation of $\beta_\text{eff}$ as a timescale. The important fact in each case is that, without needing to know the particular details of the nucleation (which could \textit{a posteriori} be determined from features of $\Gamma(t)$, $\nu(t_L)$, etc.), $\beta_{\nu}$ \textit{a priori} reproduces the correct timescale regardless of the transition. In addition, the pathologies present in both $\beta_1$ and $\beta_2$ do not occur. In particular, $\beta_{\nu} \geq 0$ always.

\section{Effects on Gravitational Wave Power Spectrum}
\label{sec:effects_on_gws}

The lifetime distribution defined above enters the calculation of the GW power spectrum via the sound-shell model (SSM). In the SSM, the GW power spectrum is related to the power spectrum of the perturbed fluid velocity field sourced by the nucleation and collision of vacuum bubbles. Calculating this fluid power spectrum requires an ensemble average over possible bubble configurations, which in turn requires averaging over the possible lifetimes of nucleated bubbles. This is where the (dimensionless) lifetime distribution enters, since it determines each bubble's contribution to the total fluid velocity field. Further details of this calculation can be found in \cite{Hindmarsh:2016lnk, Hindmarsh:2019phv, RoperPol:2023dzg}, and of its numerical implementation in \cite{Linton:2026fpj}. The essential point is that the acoustic GW power spectrum depends on $\nu(t_L)$; we therefore briefly investigate the effect of different lifetime distributions on the overall GW signal using \code{PhaseTracer2} together with a modified version of \code{HydroGrav} \cite{Linton:2026fpj}.

\begin{figure}[t]
    \centering
    \includegraphics[width=\linewidth]{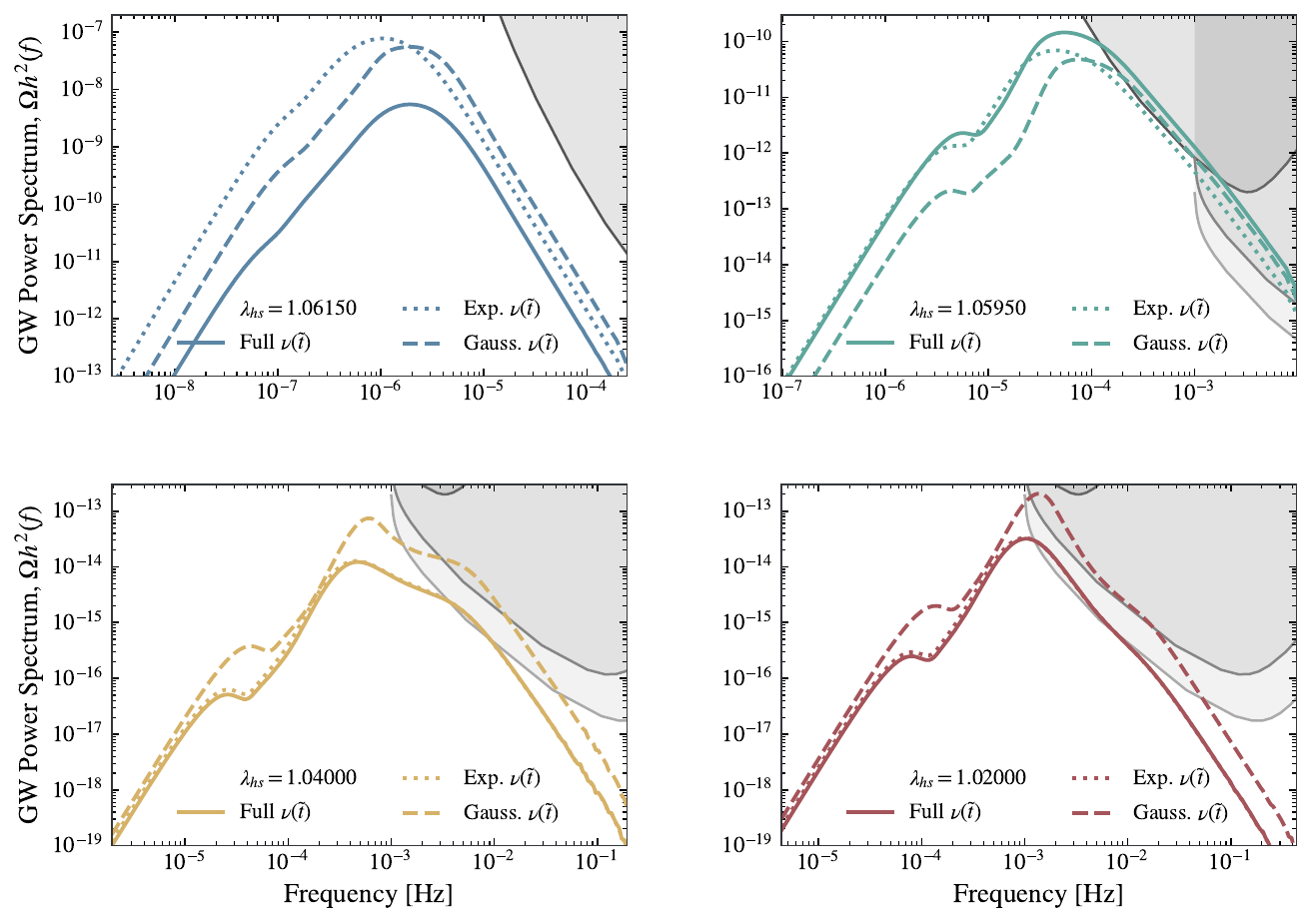}
    \caption{Comparison of GW power spectrum using different variations of the lifetime distribution. Four benchmark points of the xSM model are shown, with the exponential (dotted) and Gaussian (dashed) approximations indicated. The spectra calculated with the full lifetime distribution are shown in the solid line. The shaded regions indicate the peak-integrated sensitivities for SNR $> 10$ of LISA (darkest), DECIGO (medium), and BBO (lightest).}
    \label{fig:spectra}
\end{figure}

\code{HydroGrav} assumes either the exponential or Gaussian form of $\nu(t_L)$ during the calculation, as is conventionally done in the literature of the SSM. We can treat these cases as a benchmark, and compare them to the spectra obtained using the full calculation of $\nu(t_L)$. When evaluating $\tilde{t} = \beta t$ in each case, we will assume $\beta = \beta_1$ for exponential nucleation, and $\beta = \beta_\text{eff}$ for the Gaussian nucleation\footnote{This is provided $\beta_1 > 0$. If $\beta_1 \leq 0$, we follow convention and use $\beta_\text{eff}$ even for the exponential case. However, in such cases the nucleation will likely be Gaussian, and hence $\beta = \beta_2$ anyway.}. When we compare these benchmarks to the spectrum obtained using the full $\nu(t_L)$, we instead use $\beta_{\nu}$. This comparison is shown in \cref{fig:spectra} for four benchmark points drawn from the xSM parameter space: $\lambda_{hs} \in \{ 1.0200, 1.0400, 1.0595, 1.0615 \}$.

The spectrum for the full calculation is shown in the solid lines, whilst the exponential and Gaussian limits are the dotted and dashed lines, respectively. There is little difference in the results obtained for the strongly exponential transitions (bottom row), in which both $\nu(\tilde{t}) \approx e^{-\tilde{t}}$ and $\beta_1 \approx \beta_{\nu}$. The discrepancy between the exponential and Gaussian spectra for these plots is due to $\beta_\text{eff}$ being smaller than $\beta_1$. Although in this regime, the Gaussian spectra can be disregarded because the transition is exponential. For the slower, Gaussian transitions (top row), the calculation of the full lifetime distribution and timescale $\beta_{\nu}$ results in large discrepancies for the power spectrum. For the most Gaussian transition (top left), the peak amplitude is misrepresented by over an order of magnitude when comparing the naive exponential assumption to the full calculation. Furthermore, in the hybrid regime (top right), the long tail of the distribution introduces minor qualitative differences between the three spectra.

It may be remarked that these differences for slow transitions only occur for a very small edge case in the parameter space. However, for the xSM it is these slow transitions that yield the strongest GW signals, and hence have the best prospects for detection at a future observatory. Thus, whilst it may be the case that this occurs in a minor region of parameter space, it is exactly the region that may be observed. Notably, however, $\lambda_{hs} = 1.0615$ percolates with a super-horizon bubble separation, $R_s H > 1$. The resulting peak frequency obtained from redshifting $(k R_s)_\star$ to the present day then falls outside the observable region. 

A final remark can be made regarding the reliability of these results. The spectra shown in \cref{fig:spectra} were obtained by interfacing \code{PhaseTracer2} and \code{HydroGrav}. As such, this calculation unifies the consistent treatment of bubble nucleation, the false vacuum fraction, and timescales of this work with the consistent approach to the EoS used in \cite{Linton:2026fpj}. These spectra likely represent one of the most robust end-to-end predictions of the acoustic GW spectra for the xSM to date. Furthermore, these spectra are unique in that they were calculated without using $\alpha$ or $\beta$, in the sense that $\alpha$ is replaced by the full EoS, and $\beta$ is simply used to normalise the full lifetime distribution. In the future, improvements could be made by incorporating the calculation of the false vacuum decay rate from \code{BubbleDet} for the xSM potential, and by using a precise estimate of the wall velocity from, for example, \code{WallGo} \cite{Ekstedt:2024fyq}\footnote{In this work, we use the LTE estimate from \cite{Ai:2023see}.}.

\section{Conclusions}
\label{sec:conclusions}

In this work, we have studied bubble nucleation in cosmological phase transitions and its impact on the associated transition timescales and GW power spectrum. To this end, we first establish a self-consistent pipeline for calculating both the false vacuum decay rate and false vacuum fraction. For the decay rate, we use the full calculation of the functional determinant in the prefactor where possible, and we show that this introduces a percentage-level shift in the predicted percolation temperature relative to standard approximations. This refined decay rate calculation is paired with a novel approach for calculating the false vacuum fraction, based on solving a coupled system of ODEs. This approach avoids approximations commonly employed, such as simplifying the Hubble rate, time–temperature relationship, or EoS, by coupling the Friedmann equation for the evolution of the energy-momentum density and expansion directly to the JMAK equation for the extended volume of nucleated bubbles. This yields a self-consistent evaluation of the false vacuum fraction that incorporates all of these processes simultaneously.

We then review current methods for calculating timescales in cosmological phase transitions, focusing on the exponential and Gaussian nucleation rate approximations. After discussing the shortcomings and pathologies that arise in each case, we introduce a novel timescale, the average bubble lifetime, together with the lifetime distribution needed to calculate it. We show that the average bubble lifetime reproduces both limiting cases while remaining well-defined for edge cases and intermediate transitions that neither approximation captures. It is independent of the reference temperature, $T_\star$, and can be used a priori without any additional information about the nucleation. In the context of the GW spectrum, we argue it provides a more natural timescale and can be considered a rigorous counterpart to the effective nucleation rate, $\beta_\text{eff}$.

Finally, we propagate the full lifetime distribution and average bubble lifetime into the resulting GW power spectrum using the sound shell model. This yields an improved calculation of the GW spectrum that no longer depends on either the thermal parameter $\alpha$ or the thermal parameter $\beta$. The former is replaced by the full EoS, whilst the latter is replaced by the full lifetime distribution; $\beta_{\nu}$ appears only as a normalisation of this distribution. Using this improved calculation, we reproduce the approximation in the exponential nucleation limit. However, we identify an important region of parameter space where the improved lifetime distribution captures physics absent from both the exponential and Gaussian approximations, resulting in an order-of-magnitude discrepancy in the peak amplitude of the GW power spectrum, along with qualitative differences in the spectral features. In recent work complementary to our own, it has been shown that the GW spectrum resulting from bubble collisions in an FLRW universe is largely insensitive to the specific nucleation history \cite{Yamada:2025cfr, Yamada:2025hfs}, indicating that the dependence on bubble nucleation is source-dependent. Future work should combine our generic description of bubble nucleation with the analytic treatment of these papers to determine whether the sensitivity we find in the SSM persists for bubble collisions in less strongly supercooled transitions.

Since meaningful inference from future GW observations depends on accurate predictions of the phase transition power spectrum, we anticipate that the framework developed here, particularly the treatment of the lifetime distribution, will be relevant to improving parameter estimation pipelines for cosmological phase transitions.

\acknowledgments
W.S. is supported by the Commonwealth through an Australian Government Research Training Program Scholarship, \href{https://doi.org/10.82133/C42F-K220}{doi.org/10.82133/C42F-K220}.
C.B. is supported by Australian Research Council grants DP220100643 and LE250100010.

\appendix
\crefalias{section}{appendix}

\section{The Effective Potential}
\label{app:effective_potential}

The effective potential determines the free energy of the system through the following relationship \cite{Laine:2016hma}:
\begin{equation}
    \mathcal{F} = \veff (\phim(T), T) + \mathcal{O} \left ( \frac{ \ln \mathcal{V} }{\mathcal{V}} \right ),
\end{equation}
where $\phi_m(T)$ is the minimum of $\veff(\phi, T)$ and is in general temperature dependent. As such, thermodynamic quantities such as the energy, pressure, and entropy densities are determined from $V_\text{eff}$. Its calculation comprises a key part of the computational pipeline for cosmological phase transitions, and is one that has been heavily scrutinised \cite{Croon:2020cgk, Athron:2022jyi, Lewicki:2024xan}. In this work, we consider the one-loop effective potential:
\begin{equation}
    V_\text{eff}(\phi, T) = V_0(\phi) + V_\text{CW}(\phi) + V_{1,T}(\phi, T).
\end{equation}
Here, $V_0(\phi)$ is the tree-level contribution, whilst $V_\text{CW}$ is the zero-temperature one-loop contribution, referred to as the Coleman-Weinberg potential \cite{Coleman:1973jx}. It can be obtained by expanding the effective action to $\mathcal{O}(\hbar)$ and evaluating the resulting functional determinant. Subtracting poles in the $\ms$ scheme, it is given by\footnote{Throughout, we are assuming the Landau gauge, $\xi = 0$.}:
\begin{equation}
    V_\text{CW}(\phi) = \frac{1}{4(4 \pi)^2} \sum_{i \in \mathcal{I}} n_i (-1)^{2 s_i} m_i^4(\phi) \left [ \log \left ( \frac{m_i^2(\phi)}{\mu^2} \right) - k_i \right ].
\end{equation}
The sum $i$ contains all relevant fields, and $s_i$ is the spin of particle $i$, $n_i$ are the degrees of freedom, and $k_i$ is a number taking the value $3/2$ for scalar and fermion fields, and $5/6$ for gauge bosons. 

For the xSM, we include all SM particles in $\mathcal{I}$. The field-dependent masses for the CP-even scalars, $m_h^2$ and $m_s^2$, can be obtained by diagonalising the mass matrix:
\begin{equation}
    M_{ij}^2 = \begin{pmatrix}
        - \mu_h^2 + 3 \lambda_h \phi_h^2 + \frac{1}{2} \lambda_{hs}^2 \phi_s^2 & \lambda_{hs} \phi_h \phi_s \\ \lambda_{hs} \phi_h \phi_s & - \mu_s^2 + 3 \lambda_s \phi_s^2 + \frac{1}{2} \lambda_{hs}^2 \phi_h^2
    \end{pmatrix},
\end{equation}
whilst Goldstone scalars have masses:
\begin{equation}
    m_{G^0}^2 = m_{G^\pm}^2 = - \mu_h^2 + \lambda_h \phi_h^2 + \frac{1}{2} \lambda_{hs} \phi_s^2.
\end{equation}
For gauge boson masses:
\begin{equation}
    \quad m_W^2 = \frac{1}{4} g_W^2 \phi_h^2, \quad m_Z^2 = \frac{1}{4}(g_W^2 + g_Y^2) \phi_h^2, \quad m_A^2 = m_G^2 = 0,
\end{equation}
and all fermion masses have the form $m_{f}^2 = y_f^2 \phi_h^2 / 2$. In all cases we have suppressed the dependence on the fields $(\phi_h, \phi_s)$. For the degrees of freedom, all scalar fields have $n_s = 1$, whilst transverse and longitudinal vector fields have $n_{V} = 2$ and $n_{V} = 1$, respectively. For the fermions, quarks have $n_f = 12$, charged leptons have $n_f = 4$, and neutrinos have $n_f =2$. All couplings have been obtained from the PDG \cite{ParticleDataGroup:2024cfk}.

In the conformal $U(1)$ model, this sum is performed over the fields $\mathcal{I} = \{\phi, \varphi, A', \chi_1, \chi_2 \}$, with field-dependent masses:
\begin{equation}
    m_\phi^2(\phi) = 3 \lambda \phi_b^2, \quad m_\varphi^2(\phi_b) = \lambda \phi_b^2, \quad m_{A'}^2(\phi_b) = g^2 \phi_b^2, \quad m_{\chi_i}^2(\phi_b) = \frac{y^2 \phi_b^2}{2}.
\end{equation}
For the degrees of freedom, we have:
\begin{equation}
    n_\phi = n_\varphi = 1, \quad n_{\chi_i} = 2, \quad n_{A'} = 3.
\end{equation}

In addition to the zero-temperature contribution, interactions between the scalar fields and the thermal plasma give rise to a finite-temperature contribution to the effective potential \cite{Dolan:1973qd}. This has the following form:
\begin{equation}
    V_{1T}(\phi, T) = \frac{T^4}{2 \pi^2} \sum_i n_i (-1)^{s_i} J_{B/F} \left (\frac{m_i^2(\phi)}{T^2} \right ),
\end{equation}
where $J_{B/F}(y^2)$ is the bosonic/fermionic thermal function, given by:
\begin{equation}
    J_{B/F}(y^2) = \int_0^{\infty} d y \ y^2 \log \left ( 1 \mp e^{- \sqrt{k^2 + y^2}} \right ),
\end{equation}
where the minus(plus) sign is taken for boson(fermion) fields. For high-temperatures, $m \ll T$, these functions can be expanded to give:
\begin{equation}
    J_B(y^2) = - \frac{\pi^4}{45} + \mathcal{O}(y^2),
\end{equation}
and similarly for $J_F(y^2)$. This then provides the field-independent contribution to the energy and pressure densities, $\rho \sim T^4$. Similarly, a low-temperature expansion, $m \gg T$, yields Boltzmann suppression, $\rho \sim e^{-m/T}$. This is why all fields have been included in the xSM case, even though many contribute negligibly to the CW potential. To correctly account for their effect on the thermal bath, these fields must be included in the temperature-dependent term. 


Lastly, to deal with infrared divergences at finite temperature, we employ daisy resummation using the Parwani method \cite{Parwani:1991gq}. At one-loop order, this is equivalent to replacing all masses in the effective potential with their thermal counterparts, $m_i^2 \rightarrow m_i^2 + \Pi_i(T)$, where $i$ only runs over bosonic fields. In the xSM case, these thermal masses are given by:
\begin{align}
    \Pi_h(T) & = \left ( 9 g_W^2 + 3 g_Y^2 + 12 y_t^2 + 12 y_b^2 + 4 y_\tau^2 + 24 \lambda_h + 2 \lambda_s \right) \frac{T^2}{48}, \\
    \Pi_s(T) & = \left ( 2 \lambda_{hs} + 3 \lambda_s \right) \frac{T^2}{12}, \\
    \Pi_W(T) & = \frac{11}{6} g_W^2 T^2, \\
    \Pi_{ab}(T) & = \frac{11}{6} \begin{pmatrix}
        g_W^2 & 0 \\ 0 & g_Y^2
    \end{pmatrix} T^2, \\
    \Pi_{G}(T) & = 2 g_S^2 T^2,
\end{align}
where $\Pi_{ab}(T)$ is added to the $(W^3, B)$ mass matrix before diagonalising, and $\Pi_{G^0}(T) = \Pi_{G^\pm}(T) = \Pi_h(T)$. In the conformal $U(1)$ model, we have:
\begin{equation}
    \Pi_\phi(T) = \Pi_\varphi(T) = \left ( \frac{\lambda}{3} + \frac{y^2}{12} + \frac{g^2}{4} \right) T^2, \quad \Pi_{A'}(T) = \frac{5}{12} g^2 T^2.
\end{equation}

\section{Friedmann-JMAK Equations}
\label{app:friedmann-jmak_derivation}

In \cref{sec:false_vacuum_frac}, we provide the coupled Friedmann-JMAK equations governing a cosmological phase transition. Here, we provide a derivation of this result. We assume the universe is well modelled by a single fluid component with energy density $\bar{\rho}$, obtained by taking the weighted contributions from both the true and false vacuums, respectively:
\begin{equation}
    \bar{\rho} = h\rho_f + f \rho_t = (1-f) \rho_f + f \rho_t,
\end{equation}
where $f$ and $h$ are the true and false vacuum fractions, respectively. The energy density in each phase can be obtained from the effective potential, $V_\text{eff}(\phi, T)$, evaluated in the minima corresponding to the true and false vacuums:
\begin{equation}
    p_i(T) = - V(\phi_\text{min}(T), T), \quad \rho_i(T) = V(\phi_\text{min}(T), T) - T \frac{d V(\phi_\text{min,i}(T),T)}{dT},
    \label{eq:equation_of_state}
\end{equation}
where $i = t, f$. The exact form of the effective potential depends on the method used to construct it, and we detail our potential in \cref{app:effective_potential}. Here, we can remain agnostic to the exact structure of the potential and instead assume it is made up of both temperature-dependent and independent parts, together with a normalisation term:
\begin{equation}
    V_\text{eff}(\phi, T) = V_{T=0}(\phi) + V_T(\phi, T) + \Lambda.
\end{equation}
We normalise the potential such that $\Lambda = - V_\text{eff}(\phi_t(0), 0)$, where it is assumed $\phi_t$ represents the zero-temperature ground state. This normalises away any vacuum-energy contributions in the ground state, but not in the false vacuum. Consequently, these equations can lead to temporary periods of vacuum domination during sufficiently supercooled phase transitions.

We similarly define the average pressure density $\bar{p} = h p_f + f p_t$. Together, the Friedmann and conservation equations are then given by:
\begin{equation}
    H^2 = \frac{8 \pi G}{3} \bar{\rho}, \quad \dot{\bar{\rho}} = - 3 H (\bar{\rho} + \bar{p}).
    \label{eq:hubble_and_consv_eq}
\end{equation}
These equations evolve the averaged energy and pressure densities. To isolate $\rho_i$ and $p_i$, we assume the false vacuum evolves independently:
\begin{equation}
    \dot{\rho}_f = - 3 H (\rho_f + p_f).
    \label{eq:rho_f_dot}
\end{equation}
This is akin to assuming no energy transfer between the true and false vacuums in front of the bubble wall \cite{Matuszak:2026xsz}. Inserting this into the continuity equation allows us to write:
\begin{equation}
    \dot{\rho}_t = - 3 H (\rho_t + p_t) + \frac{\dot{f}}{f} (\rho_f - \rho_t) 
    \label{eq:rho_t_dot}.
\end{equation}
The second term in \cref{eq:rho_t_dot} represents the effect of reheating: latent heat from the decaying false vacuum is injected into the interior plasma within the true vacuum bubble. This manifests as an increased temperature of the true vacuum, $T_t > T_f$.

An equation of state is needed to break the degeneracy between $\rho$ and $p$, which we obtain from the effective potential \cref{eq:equation_of_state}. As such, the pressure is not an independent degree of freedom. Given $\rho_i(t)$, the corresponding temperature of each phase can be determined by matching $\rho_i(t) = \rho(T_i)$, where the term on the RHS is the energy density as defined in \cref{eq:equation_of_state}. Once the temperature is known, the corresponding pressure can be found from the equation of state $p_i(t) = p(T_i)$, where again $p(T_i)$ is given by \cref{eq:equation_of_state}. Note that in the false vacuum fraction, this procedure is equivalent to evolving $T_f(t)$ according to:
\begin{equation}
    \frac{d T_f}{dt} - 3 H \frac{\rho_f + p_f}{d \rho_f / dT},
\end{equation}
obtained by applying the chain rule to \cref{eq:rho_f_dot}. Unfortunately, no such time-temperature relationship can be used for the true vacuum, and the method outlined above must be used to implicitly determine $T_t$. In any case, the solutions of \cref{eq:rho_f_dot} and \cref{eq:rho_t_dot}, together with the equation of state \cref{eq:equation_of_state}, uniquely determine $\{\rho_t, \rho_f, p_t, p_f, T_t, T_f \}$ at each time.

The independent degrees of freedom have been reduced to the Hubble rate, $H(t)$, each energy density, $\rho_t(f)$ and $\rho_f(t)$, and the true vacuum fraction, $f(t)$; governed by three constraints: the first Friedmann equation \cref{eq:hubble_and_consv_eq}, together with the conservation equations \cref{eq:rho_f_dot} and \cref{eq:rho_t_dot}. To fully determine the system, we must evolve either the true or false vacuum fraction. It will be more convenient to evolve the latter, which is defined by:
\begin{equation}
    h(t) = e^{-\mathcal{V}_\text{ext}(t)},
    \label{eq:false_vacuum_fraction_def}
\end{equation}
where $\mathcal{V}_\text{ext}$ is the extended volume of nucleated bubbles:
\begin{equation}
    \mathcal{V}_\text{ext}(t) = \int_{t_0}^{t} dt' \Gamma(t') V(t', t) \left ( \frac{a(t')}{a(t)} \right )^3.
\end{equation}
Here, $V(t', t) = \frac{4 \pi}{3} R(t', t)^3$ is the physical volume of bubbles nucleated at time $t'$ at some later time $t$. In general, the bubble radius $R(t', t)$ is given by:
\begin{equation}
    R(t', t) = \frac{a(t)}{a(t')} R_0(t') + \int_{t'}^t dt'' v_w(t', t'') \frac{a(t)}{a(t'')}.
\end{equation}
We simplify this expression with two assumptions; first, we assume the initial radius is negligible $R_0 \sim 0$. This initial radius can be determined from the bounce solution, and is typically of the order $T^{-1}$ \cite{Megevand:2007sv}. It follows $R_0 / \delta t \sim R_0 H \sim T / M_\text{pl} \ll 1$. Secondly, we assume the bubble wall rapidly accelerates to a constant velocity, $v_w$. With these, we may write the false vacuum fraction as:
\begin{equation}
   h(t) = \exp \left ( - \frac{4 \pi v_w^3}{3} I_3(t) \right ),
   \label{eq:false_vac_with_I3}
\end{equation}
where the integral $I_n$ is given by:
\begin{equation}
    I_n(t) = \int_{t_0}^t dt' \Gamma(t') \left ( \frac{a(t')}{a(t)} \right)^3 \left ( \int_{t'}^t dt'' \frac{a(t)}{a(t'')} \right )^n = \int_{t_0}^t dt' \Gamma(t') \left ( \frac{a(t')}{a(t)} \right)^3 K(t', t)^n.
\end{equation}
The difficulty in the false vacuum fraction calculation should now be apparent. The false vacuum fraction depends on the scale factor $a(t)$, and so implicitly depends on the solution to $H(t)$. However, determining $H(t)$ requires the false vacuum fraction. Furthermore, the difficulty is compounded by the fact that the expression involves a series of nested integrals\footnote{Writing $a(t_1)/a(t_2) = \exp(\int dt' H(t))$ demonstrates this requires a series of three nested integrals.}. Fortunately, we employ a trick utilised in \cite{Matuszak:2026xsz} to simplify this problem extensively. The integral $I_n(t)$ can be differentiated as a function of $t$, taking care to apply the Leibniz rule. Each integration reduces the index on $K(t', t)$, until after $n$ integrations, the final innermost integral is removed and the expression is instead reduced to a system of ordinary differential equations, satisfying the recursive relationship:
\begin{equation}
    \dot{I}_n(t) = n I_{n-1}(t) + (n-3)H(t) I_n(t).
    \label{eq:I_n_deriv}
\end{equation}
Differentiating $I_0$ yields the initial condition:
\begin{equation}
    \dot{I}_0(t) = \Gamma(t) - 3 H(t) I_0(t).
    \label{eq:I_0_deriv}
\end{equation}
As such, the role of $f(t)$ has been replaced with the coupled set of four ordinary differential equations given by \cref{eq:I_n_deriv} and \cref{eq:I_0_deriv}, where, once $I_3(t)$ is known, the false vacuum fraction can be determined using \cref{eq:false_vac_with_I3}, and the true vacuum fraction from $f(t) = 1 - h(t)$.

The system of equations, including the Friedmann equation \cref{eq:hubble_and_consv_eq}, conservation equations \cref{eq:rho_f_dot} and \cref{eq:rho_t_dot}, as well as the equations for $I_n$, \cref{eq:I_n_deriv} and \cref{eq:I_0_deriv} thus form a completely determined system for the quantities $\{H, \rho_t, \rho_f, f \}$. All other relevant metrics can then be derived from this set. Within the main body of text, we show the solution for these equations for two benchmark points.

\section{Bubble Lifetime Distribution}
\label{app:lifetime_distribution}

In this appendix, we provide a novel derivation of the lifetime distribution, \cref{eq:lifetime_dist_full}. Recall $\nu(t_L)$ is defined such that the probability a bubble disappears with a lifetime inside the small interval $[t_L, t_L + dt_L]$ is given by $\nu(t_L) dt_L$. It follows:
\begin{equation}
    \int_{0}^\infty dt_L \nu (t_L) = 1.
\end{equation}
We take a bubble to disappear once its nucleation point is engulfed by the wall of another bubble, consistent with our discussion of $\beta_\text{eff}$ and with \cite{Hindmarsh:2019phv}. This is slightly later than the time of first contact between walls, with the time difference being of the order of $R/v_w$, where $R$ is the bubble radius.

To proceed, we write $\nu(t_L)$ in the form:
\begin{equation}
    \nu(t_L) = \frac{1}{\mathcal{N}_\infty} \frac{d \mathcal{N}}{d t_L},
\end{equation}
where we have introduced the comoving bubble density $\mathcal{N}(t) = n(t) a(t)^3$, where $n(t)$ is given by \cref{eq:mean_bubble_density}. $\mathcal{N}_\infty = \mathcal{N}(t \rightarrow \infty)$ is the total comoving number density.
where $n_\infty = n(t \rightarrow \infty)$ is the total number density of nucleated bubbles. The second term can be written as the marginalised probability distribution:
\begin{equation}
    \frac{d \mathcal{N}}{d t_L} = \int_{t_0}^\infty d t_n P(t_L | t_n) \frac{d \mathcal{N}}{d t_n},
    \label{eq:lifetime_dist_inter}
\end{equation}
where $P(t_L | t_n)$ is the conditional probability that a bubble nucleated at $t_n$ survives with a lifetime $t_L$. Written this way, we interpret the lifetime distribution as follows:
\begin{equation}
    \nu(t_L) = \int_{t_0}^\infty P(\text{lifetime } t_L | \text{nucleates at } t_n) P(\text{nucleates } t_n).
\end{equation}

The second term in the integrand is obtained from the definition of the comoving number density:
\begin{equation}
    \mathcal{N}(t_n) = n(t_n)a(t_n)^3 = \int_{t_0}^{t_n} d t' \Gamma(t') h(t') a(t')^3,
\end{equation}
such that upon differentiating, we have:
\begin{equation}
    \frac{d \mathcal{N}}{dt_n} = \Gamma(t_n) h(t_n) a(t_n)^3.
    \label{eq:dN_dtn}
\end{equation}
When compared to, e.g., \cref{eq:I_0_deriv}, the dilution term $\propto -3H$ is absent, being absorbed into the comoving nature of $\mathcal{N}$.

The first term $P(t_L \,|\, t_n)$ is evaluated as a hazard rate problem. Let $A_t$ denote the event that a given co-moving point lies in the false vacuum at time $t$, so that $P(A_t) = h(t)$. Then, define the survival probability, $S(t \,|\, t')$, as the probability that a bubble nucleated at $t'$ is still present at $t$, and the hazard rate, $\lambda(t)$, as the probability per unit time that a surviving bubble is destroyed at $t$. It follows
\begin{equation}
    P(t_L \,|\, t_n) = \lambda(t_n + t_L)\, S(t_n + t_L \,|\, t_n).
    \label{eq:conditional_def}
\end{equation}
The hazard rate follows from the rate at which space is converted from the false to the true vacuum,
\begin{equation}
    \lambda(t) = -\frac{1}{h(t)} \frac{dh}{dt},
\end{equation}
while the survival probability is
\begin{equation}
    S(t_n + t_L \,|\, t_n)
      = P\!\left( A_{t_n + t_L} \,|\, A_{t_n} \right)
      = \frac{P\!\left( A_{t_n + t_L} \cap A_{t_n} \right)}{P\!\left( A_{t_n} \right)}
      = \frac{P\!\left( A_{t_n + t_L} \right)}{P\!\left( A_{t_n} \right)}
      = \frac{h(t_n + t_L)}{h(t_n)}.
\end{equation}
The second equality is the definition of conditional probability, and the third follows because $A_{t_n + t_L} \subseteq A_{t_n}$. That is, any point in the false vacuum at $t_n + t_L$ was necessarily in the false vacuum at $t_n$. 

Combining these results, we find the conditional probability \cref{eq:conditional_def} is:
\begin{equation}
    P(t_L | t_n) = - \frac{1}{h(t_n)} \frac{d h}{d t'} \Bigg |_{t' = t_n + t_L}.
    \label{eq:conditional_prob_evaluated}
\end{equation}
Lastly, using both \cref{eq:dN_dtn} and \cref{eq:conditional_prob_evaluated} in \cref{eq:lifetime_dist_inter} gives the final result:
\begin{equation}
    \nu(t_L) = - \frac{1}{\mathcal{N}_\infty} \int_{t_0}^\infty d t_n \Gamma(t_n) a(t_n)^3\frac{dh}{dt'} \Bigg |_{t' = t_n + t_L}.
\end{equation}
This is consistent with the expression obtained in \cite{Hindmarsh:2019phv}, although the above result generalises to expanding spacetime. In \cite{Hindmarsh:2019phv}, it is shown that if we neglect the expansion of spacetime, $a(t) \rightarrow 1$ and $\mathcal{N} \rightarrow n$, then $\nu(t)$ can be evaluated analytically for exponential and Gaussian nucleation. Writing the dimensionless lifetime distribution as a function of $\tilde{t} = \beta t_L$, we have \cite{Hindmarsh:2019phv}:
\begin{equation}
    \nu_\text{exp}(\tilde{t}) = e^{-\tilde{t}}, \quad \nu_\text{sim}(\tilde{t}) = \frac{1}{2} \tilde{t}^2 e^{-\frac{1}{6}\tilde{t}^3}.
    \label{eq:nu_approx_app}
\end{equation}
These are the asymptotic forms for $\nu(\tilde{t})$ used in the text.

\bibliographystyle{JHEP}
\bibliography{biblio.bib}

@article{RoperPol:2023dzg,
    author = "Roper Pol, Alberto and Procacci, Simona and Caprini, Chiara",
    title = "{Characterization of the gravitational wave spectrum from sound waves within the sound shell model}",
    eprint = "2308.12943",
    archivePrefix = "arXiv",
    primaryClass = "gr-qc",
    doi = "10.1103/PhysRevD.109.063531",
    journal = "Phys. Rev. D",
    volume = "109",
    number = "6",
    pages = "063531",
    year = "2024"
}

@article{Hindmarsh:2016lnk,
    author = "Hindmarsh, Mark",
    title = "{Sound shell model for acoustic gravitational wave production at a first-order phase transition in the early Universe}",
    eprint = "1608.04735",
    archivePrefix = "arXiv",
    primaryClass = "astro-ph.CO",
    doi = "10.1103/PhysRevLett.120.071301",
    journal = "Phys. Rev. Lett.",
    volume = "120",
    number = "7",
    pages = "071301",
    year = "2018"
}

@article{Hindmarsh:2019phv,
    author = "Hindmarsh, Mark and Hijazi, Mulham",
    title = "{Gravitational waves from first order cosmological phase transitions in the Sound Shell Model}",
    eprint = "1909.10040",
    archivePrefix = "arXiv",
    primaryClass = "astro-ph.CO",
    reportNumber = "NORDITA-2019-083, HIP-2019-29/TH",
    doi = "10.1088/1475-7516/2019/12/062",
    journal = "JCAP",
    volume = "12",
    pages = "062",
    year = "2019"
}

@article{Athron:2024xrh,
    author = "Athron, Peter and Balazs, Csaba and Fowlie, Andrew and Morris, Lachlan and Searle, William and Xiao, Yang and Zhang, Yang",
    title = "{PhaseTracer2: from the effective potential to gravitational waves}",
    eprint = "2412.04881",
    archivePrefix = "arXiv",
    primaryClass = "astro-ph.CO",
    doi = "10.1140/epjc/s10052-025-14258-y",
    journal = "Eur. Phys. J. C",
    volume = "85",
    number = "5",
    pages = "559",
    year = "2025"
}

@article{Ekstedt:2023sqc,
    author = "Ekstedt, Andreas and Gould, Oliver and Hirvonen, Joonas",
    title = "{BubbleDet: a Python package to compute functional determinants for bubble nucleation}",
    eprint = "2308.15652",
    archivePrefix = "arXiv",
    primaryClass = "hep-ph",
    doi = "10.1007/JHEP12(2023)056",
    journal = "JHEP",
    volume = "12",
    pages = "056",
    year = "2023"
}

@article{Wainwright:2011kj,
    author = "Wainwright, Carroll L.",
    title = "{CosmoTransitions: Computing Cosmological Phase Transition Temperatures and Bubble Profiles with Multiple Fields}",
    eprint = "1109.4189",
    archivePrefix = "arXiv",
    primaryClass = "hep-ph",
    doi = "10.1016/j.cpc.2012.04.004",
    journal = "Comput. Phys. Commun.",
    volume = "183",
    pages = "2006--2013",
    year = "2012"
}

@article{Athron:2023rfq,
    author = "Athron, Peter and Morris, Lachlan and Xu, Zhongxiu",
    title = "{How robust are gravitational wave predictions from cosmological phase transitions?}",
    eprint = "2309.05474",
    archivePrefix = "arXiv",
    primaryClass = "hep-ph",
    doi = "10.1088/1475-7516/2024/05/075",
    journal = "JCAP",
    volume = "05",
    pages = "075",
    year = "2024"
}

@article{Hindmarsh:2015qta,
    author = "Hindmarsh, Mark and Huber, Stephan J. and Rummukainen, Kari and Weir, David J.",
    title = "{Numerical simulations of acoustically generated gravitational waves at a first order phase transition}",
    eprint = "1504.03291",
    archivePrefix = "arXiv",
    primaryClass = "astro-ph.CO",
    reportNumber = "HIP-2015-13-TH",
    doi = "10.1103/PhysRevD.92.123009",
    journal = "Phys. Rev. D",
    volume = "92",
    number = "12",
    pages = "123009",
    year = "2015"
}

@article{Hindmarsh:2017gnf,
    author = "Hindmarsh, Mark and Huber, Stephan J. and Rummukainen, Kari and Weir, David J.",
    title = "{Shape of the acoustic gravitational wave power spectrum from a first order phase transition}",
    eprint = "1704.05871",
    archivePrefix = "arXiv",
    primaryClass = "astro-ph.CO",
    reportNumber = "HIP-2017-02-TH, HIP-2017-02/TH",
    doi = "10.1103/PhysRevD.96.103520",
    journal = "Phys. Rev. D",
    volume = "96",
    number = "10",
    pages = "103520",
    year = "2017",
    note = "[Erratum: Phys.Rev.D 101, 089902 (2020)]"
}

@article{Chandrasekhar:1943ws,
    author = "Chandrasekhar, Subrahmanyan",
    editor = "Wali, K. C.",
    title = "{Stochastic problems in physics and astronomy}",
    doi = "10.1103/RevModPhys.15.1",
    journal = "Rev. Mod. Phys.",
    volume = "15",
    pages = "1--89",
    year = "1943"
}

@article{Matuszak:2026xsz,
    author = "Matuszak, Jonas and Tasillo, Carlo",
    title = "{TransitionListener v2.0 -- Robust gravitational wave predictions for cosmological phase transitions}",
    eprint = "2605.15259",
    archivePrefix = "arXiv",
    primaryClass = "hep-ph",
    month = "5",
    year = "2026"
}

@article{Callan:1977pt,
    author = "Callan, Jr., Curtis G. and Coleman, Sidney R.",
    title = "{The Fate of the False Vacuum. 2. First Quantum Corrections}",
    reportNumber = "HUTP-77-A032",
    doi = "10.1103/PhysRevD.16.1762",
    journal = "Phys. Rev. D",
    volume = "16",
    pages = "1762--1768",
    year = "1977"
}

@article{Coleman:1977py,
    author = "Coleman, Sidney R.",
    title = "{The Fate of the False Vacuum. 1. Semiclassical Theory}",
    reportNumber = "HUTP-77-A004",
    doi = "10.1103/PhysRevD.15.2929",
    journal = "Phys. Rev. D",
    volume = "15",
    pages = "2929--2936",
    year = "1977",
    note = "[Erratum: Phys.Rev.D 16, 1248 (1977)]"
}

@article{Linde:1980tt,
    author = "Linde, Andrei D.",
    title = "{Fate of the False Vacuum at Finite Temperature: Theory and Applications}",
    reportNumber = "LEBEDEV-80-92",
    doi = "10.1016/0370-2693(81)90281-1",
    journal = "Phys. Lett. B",
    volume = "100",
    pages = "37--40",
    year = "1981"
}

@article{Linde:1977mm,
    author = "Linde, Andrei D.",
    title = "{On the Vacuum Instability and the Higgs Meson Mass}",
    reportNumber = "LEBEDEV-77-112",
    doi = "10.1016/0370-2693(77)90664-5",
    journal = "Phys. Lett. B",
    volume = "70",
    pages = "306--308",
    year = "1977"
}

@article{Masoumi:2017trx,
    author = "Masoumi, Ali and Olum, Ken D. and Wachter, Jeremy M.",
    title = "{Approximating tunneling rates in multi-dimensional field spaces}",
    eprint = "1702.00356",
    archivePrefix = "arXiv",
    primaryClass = "gr-qc",
    doi = "10.1088/1475-7516/2017/10/022",
    journal = "JCAP",
    volume = "10",
    pages = "022",
    year = "2017",
    note = "[Erratum: JCAP 05, E01 (2023)]"
}

@article{Guada:2020xnz,
    author = "Guada, Victor and Nemev{\v{s}}ek, Miha and Pintar, Matev{\v{z}}",
    title = "{FindBounce: Package for multi-field bounce actions}",
    eprint = "2002.00881",
    archivePrefix = "arXiv",
    primaryClass = "hep-ph",
    doi = "10.1016/j.cpc.2020.107480",
    journal = "Comput. Phys. Commun.",
    volume = "256",
    pages = "107480",
    year = "2020"
}

@article{Athron:2019nbd,
    author = "Athron, Peter and Bal{\'a}zs, Csaba and Bardsley, Michael and Fowlie, Andrew and Harries, Dylan and White, Graham",
    title = "{BubbleProfiler: finding the field profile and action for cosmological phase transitions}",
    eprint = "1901.03714",
    archivePrefix = "arXiv",
    primaryClass = "hep-ph",
    reportNumber = "CoEPP-MN-19-1",
    doi = "10.1016/j.cpc.2019.05.017",
    journal = "Comput. Phys. Commun.",
    volume = "244",
    pages = "448--468",
    year = "2019"
}

@article{Sato:2019wpo,
    author = "Sato, Ryosuke",
    title = "{SimpleBounce : a simple package for the false vacuum decay}",
    eprint = "1908.10868",
    archivePrefix = "arXiv",
    primaryClass = "hep-ph",
    reportNumber = "DESY 19-148, DESY-19-148",
    doi = "10.1016/j.cpc.2020.107566",
    journal = "Comput. Phys. Commun.",
    volume = "258",
    pages = "107566",
    year = "2021",
    note = "[Erratum: Comput.Phys.Commun. 320, 110006 (2026)]"
}

@article{Basler:2024aaf,
    author = {Basler, Philipp and Biermann, Lisa and M{\"u}hlleitner, Margarete and M{\"u}ller, Jonas and Santos, Rui and Viana, Jo{\~a}o},
    title = "{BSMPT v3 a tool for phase transitions and primordial gravitational waves in extended Higgs sectors}",
    eprint = "2404.19037",
    archivePrefix = "arXiv",
    primaryClass = "hep-ph",
    reportNumber = "KA-TP-08-2024",
    doi = "10.1016/j.cpc.2025.109766",
    journal = "Comput. Phys. Commun.",
    volume = "316",
    pages = "109766",
    year = "2025"
}

@article{Costa:2025pew,
    author = "Costa, Francesco and Hoefken Zink, Jaime and Lucente, Michele and Pascoli, Silvia and Rosauro-Alcaraz, Salvador",
    title = "{ELENA: a software for fast and precise computation of first order phase transitions and gravitational waves production in particle physics models}",
    eprint = "2510.00289",
    archivePrefix = "arXiv",
    primaryClass = "hep-ph",
    doi = "10.1140/epjc/s10052-026-16003-5",
    journal = "Eur. Phys. J. C",
    volume = "86",
    number = "7",
    pages = "851",
    year = "2026"
}

@article{Linton:2026fpj,
    author = "Linton, Flynn and Searle, William and Wang, Xiao and Bal{\'a}zs, Csaba",
    title = "{HydroGrav: Precise hydrodynamics and gravitational waves for cosmological phase transitions}",
    eprint = "2606.27775",
    archivePrefix = "arXiv",
    primaryClass = "hep-ph",
    month = "6",
    year = "2026"
}

@article{Croon:2020cgk,
    author = "Croon, Djuna and Gould, Oliver and Schicho, Philipp and Tenkanen, Tuomas V. I. and White, Graham",
    title = "{Theoretical uncertainties for cosmological first-order phase transitions}",
    eprint = "2009.10080",
    archivePrefix = "arXiv",
    primaryClass = "hep-ph",
    reportNumber = "HIP-2020-26/TH",
    doi = "10.1007/JHEP04(2021)055",
    journal = "JHEP",
    volume = "04",
    pages = "055",
    year = "2021"
}

@article{Athron:2022jyi,
    author = "Athron, Peter and Balazs, Csaba and Fowlie, Andrew and Morris, Lachlan and White, Graham and Zhang, Yang",
    title = "{How arbitrary are perturbative calculations of the electroweak phase transition?}",
    eprint = "2208.01319",
    archivePrefix = "arXiv",
    primaryClass = "hep-ph",
    doi = "10.1007/JHEP01(2023)050",
    journal = "JHEP",
    volume = "01",
    pages = "050",
    year = "2023"
}

@article{Lewicki:2024xan,
    author = "Lewicki, Marek and Merchand, Marco and Sagunski, Laura and Schicho, Philipp and Schmitt, Daniel",
    title = "{Impact of theoretical uncertainties on model parameter reconstruction from GW signals sourced by cosmological phase transitions}",
    eprint = "2403.03769",
    archivePrefix = "arXiv",
    primaryClass = "hep-ph",
    doi = "10.1103/PhysRevD.110.023538",
    journal = "Phys. Rev. D",
    volume = "110",
    number = "2",
    pages = "023538",
    year = "2024"
}

@article{Ai:2023see,
    author = "Ai, Wen-Yuan and Laurent, Benoit and van de Vis, Jorinde",
    title = "{Model-independent bubble wall velocities in local thermal equilibrium}",
    eprint = "2303.10171",
    archivePrefix = "arXiv",
    primaryClass = "astro-ph.CO",
    reportNumber = "KCL-PH-TH/2023-19",
    doi = "10.1088/1475-7516/2023/07/002",
    journal = "JCAP",
    volume = "07",
    pages = "002",
    year = "2023"
}

@article{Ekstedt:2024fyq,
    author = "Ekstedt, Andreas and Gould, Oliver and Hirvonen, Joonas and Laurent, Benoit and Niemi, Lauri and Schicho, Philipp and van de Vis, Jorinde",
    title = "{How fast does the WallGo? A package for computing wall velocities in first-order phase transitions}",
    eprint = "2411.04970",
    archivePrefix = "arXiv",
    primaryClass = "hep-ph",
    reportNumber = "CERN-TH-2024-174, DESY-24-162, HIP-2024-21/TH",
    doi = "10.1007/JHEP04(2025)101",
    journal = "JHEP",
    volume = "04",
    pages = "101",
    year = "2025"
}

@article{Choi:1993cv,
    author = "Choi, Jarny and Volkas, R. R.",
    title = "{Real Higgs singlet and the electroweak phase transition in the Standard Model}",
    eprint = "hep-ph/9308234",
    archivePrefix = "arXiv",
    reportNumber = "UM-P-93-80, OZ-93-20",
    doi = "10.1016/0370-2693(93)91013-D",
    journal = "Phys. Lett. B",
    volume = "317",
    pages = "385--391",
    year = "1993"
}

@article{Ham:2004cf,
    author = "Ham, S. W. and Jeong, Y. S. and Oh, S. K.",
    title = "{Electroweak phase transition in an extension of the standard model with a real Higgs singlet}",
    eprint = "hep-ph/0411352",
    archivePrefix = "arXiv",
    doi = "10.1088/0954-3899/31/8/017",
    journal = "J. Phys. G",
    volume = "31",
    number = "8",
    pages = "857--871",
    year = "2005"
}

@article{Balan:2025uke,
    author = "Balan, Sowmiya and Bringmann, Torsten and Kahlhoefer, Felix and Matuszak, Jonas and Tasillo, Carlo",
    title = "{Sub-GeV dark matter and nano-Hertz gravitational waves from a classically conformal dark sector}",
    eprint = "2502.19478",
    archivePrefix = "arXiv",
    primaryClass = "hep-ph",
    doi = "10.1088/1475-7516/2025/08/062",
    journal = "JCAP",
    volume = "08",
    pages = "062",
    year = "2025"
}

@book{Laine:2016hma,
    author = "Laine, Mikko and Vuorinen, Aleksi",
    title = "{Basics of Thermal Field Theory}",
    eprint = "1701.01554",
    archivePrefix = "arXiv",
    primaryClass = "hep-ph",
    doi = "10.1007/978-3-319-31933-9",
    publisher = "Springer",
    volume = "925",
    year = "2016"
}

@article{Coleman:1973jx,
    author = "Coleman, Sidney R. and Weinberg, Erick J.",
    title = "{Radiative Corrections as the Origin of Spontaneous Symmetry Breaking}",
    doi = "10.1103/PhysRevD.7.1888",
    journal = "Phys. Rev. D",
    volume = "7",
    pages = "1888--1910",
    year = "1973"
}

@article{Dolan:1973qd,
    author = "Dolan, L. and Jackiw, R.",
    title = "{Symmetry Behavior at Finite Temperature}",
    reportNumber = "MIT-CTP-406",
    doi = "10.1103/PhysRevD.9.3320",
    journal = "Phys. Rev. D",
    volume = "9",
    pages = "3320--3341",
    year = "1974"
}

@article{ParticleDataGroup:2024cfk,
    author = "Navas, S. and others",
    collaboration = "Particle Data Group",
    title = "{Review of particle physics}",
    doi = "10.1103/PhysRevD.110.030001",
    journal = "Phys. Rev. D",
    volume = "110",
    number = "3",
    pages = "030001",
    year = "2024"
}

@article{Parwani:1991gq,
    author = "Parwani, Rajesh R.",
    title = "{Resummation in a hot scalar field theory}",
    eprint = "hep-ph/9204216",
    archivePrefix = "arXiv",
    reportNumber = "ITP-SB-91-64",
    doi = "10.1103/PhysRevD.45.4695",
    journal = "Phys. Rev. D",
    volume = "45",
    pages = "4695",
    year = "1992",
    note = "[Erratum: Phys.Rev.D 48, 5965 (1993)]"
}

@article{Megevand:2007sv,
    author = "Megevand, Ariel and Sanchez, Alejandro D.",
    title = "{Supercooling and phase coexistence in cosmological phase transitions}",
    eprint = "0712.1031",
    archivePrefix = "arXiv",
    primaryClass = "hep-ph",
    doi = "10.1103/PhysRevD.77.063519",
    journal = "Phys. Rev. D",
    volume = "77",
    pages = "063519",
    year = "2008"
}

@article{Schwaller:2015tja,
    author = "Schwaller, Pedro",
    title = "{Gravitational Waves from a Dark Phase Transition}",
    eprint = "1504.07263",
    archivePrefix = "arXiv",
    primaryClass = "hep-ph",
    reportNumber = "CERN-PH-TH-2015-093",
    doi = "10.1103/PhysRevLett.115.181101",
    journal = "Phys. Rev. Lett.",
    volume = "115",
    number = "18",
    pages = "181101",
    year = "2015"
}

@article{Kamionkowski:1993fg,
    author = "Kamionkowski, Marc and Kosowsky, Arthur and Turner, Michael S.",
    title = "{Gravitational radiation from first order phase transitions}",
    eprint = "astro-ph/9310044",
    archivePrefix = "arXiv",
    reportNumber = "IASSNS-HEP-93-44, FERMILAB-PUB-93-235-A",
    doi = "10.1103/PhysRevD.49.2837",
    journal = "Phys. Rev. D",
    volume = "49",
    pages = "2837--2851",
    year = "1994"
}

@article{Kosowsky:1991ua,
    author = "Kosowsky, Arthur and Turner, Michael S. and Watkins, Richard",
    title = "{Gravitational Radiation from Colliding Vacuum Bubbles}",
    reportNumber = "FERMILAB-PUB-91-323-A",
    doi = "10.1103/PhysRevD.45.4514",
    journal = "Phys. Rev. D",
    volume = "45",
    pages = "4514--4535",
    year = "1992"
}

@article{Kosowsky:1992vn,
    author = "Kosowsky, Arthur and Turner, Michael S.",
    title = "{Gravitational radiation from colliding vacuum bubbles: envelope approximation to many bubble collisions}",
    eprint = "astro-ph/9211004",
    archivePrefix = "arXiv",
    reportNumber = "FERMILAB-PUB-92-295-A",
    doi = "10.1103/PhysRevD.47.4372",
    journal = "Phys. Rev. D",
    volume = "47",
    pages = "4372--4391",
    year = "1993"
}

@article{Jinno:2016vai,
    author = "Jinno, Ryusuke and Takimoto, Masahiro",
    title = "{Gravitational waves from bubble collisions: An analytic derivation}",
    eprint = "1605.01403",
    archivePrefix = "arXiv",
    primaryClass = "astro-ph.CO",
    reportNumber = "KEK-TH-1900",
    doi = "10.1103/PhysRevD.95.024009",
    journal = "Phys. Rev. D",
    volume = "95",
    number = "2",
    pages = "024009",
    year = "2017"
}

@article{Jinno:2017fby,
    author = "Jinno, Ryusuke and Takimoto, Masahiro",
    title = "{Gravitational waves from bubble dynamics: Beyond the Envelope}",
    eprint = "1707.03111",
    archivePrefix = "arXiv",
    primaryClass = "hep-ph",
    reportNumber = "CTPU-17-26, KEK-TH-1986",
    doi = "10.1088/1475-7516/2019/01/060",
    journal = "JCAP",
    volume = "01",
    pages = "060",
    year = "2019"
}

@article{Hindmarsh:2013xza,
    author = "Hindmarsh, Mark and Huber, Stephan J. and Rummukainen, Kari and Weir, David J.",
    title = "{Gravitational waves from the sound of a first order phase transition}",
    eprint = "1304.2433",
    archivePrefix = "arXiv",
    primaryClass = "hep-ph",
    reportNumber = "HIP-2013-07-TH",
    doi = "10.1103/PhysRevLett.112.041301",
    journal = "Phys. Rev. Lett.",
    volume = "112",
    pages = "041301",
    year = "2014"
}

@article{Caprini:2006jb,
    author = "Caprini, Chiara and Durrer, Ruth",
    title = "{Gravitational waves from stochastic relativistic sources: Primordial turbulence and magnetic fields}",
    eprint = "astro-ph/0603476",
    archivePrefix = "arXiv",
    doi = "10.1103/PhysRevD.74.063521",
    journal = "Phys. Rev. D",
    volume = "74",
    pages = "063521",
    year = "2006"
}

@article{Caprini:2009yp,
    author = "Caprini, Chiara and Durrer, Ruth and Servant, Geraldine",
    title = "{The stochastic gravitational wave background from turbulence and magnetic fields generated by a first-order phase transition}",
    eprint = "0909.0622",
    archivePrefix = "arXiv",
    primaryClass = "astro-ph.CO",
    doi = "10.1088/1475-7516/2009/12/024",
    journal = "JCAP",
    volume = "12",
    pages = "024",
    year = "2009"
}

@article{Bond:1984tga,
    author = {Bond, J. R. and Carr, B. J.},
    title = {Gravitational waves from a population of binary black holes},
    journal = {Monthly Notices of the Royal Astronomical Society},
    volume = {207},
    number = {3},
    pages = {585-609},
    year = {1984},
    month = {04},
    issn = {0035-8711},
    doi = {10.1093/mnras/207.3.585},
    url = {https://doi.org/10.1093/mnras/207.3.585},
    eprint = {https://academic.oup.com/mnras/article-pdf/207/3/585/18521578/mnras207-0585.pdf},
}

@article{Hawking:1982ga,
    author = "Hawking, S. W. and Moss, I. G. and Stewart, J. M.",
    title = "{Bubble Collisions in the Very Early Universe}",
    reportNumber = "Print-82-0180 (CAMBRIDGE)",
    doi = "10.1103/PhysRevD.26.2681",
    journal = "Phys. Rev. D",
    volume = "26",
    pages = "2681",
    year = "1982"
}

@article{Kibble:1976sj,
    author = "Kibble, T. W. B.",
    title = "{Topology of Cosmic Domains and Strings}",
    reportNumber = "ICTP/75/5",
    doi = "10.1088/0305-4470/9/8/029",
    journal = "J. Phys. A",
    volume = "9",
    pages = "1387--1398",
    year = "1976"
}

@article{Witten:1984rs,
    author = "Witten, Edward",
    title = "{Cosmic Separation of Phases}",
    reportNumber = "PRINT-84-0400 (IAS,PRINCETON)",
    doi = "10.1103/PhysRevD.30.272",
    journal = "Phys. Rev. D",
    volume = "30",
    pages = "272--285",
    year = "1984"
}

@article{Hogan:1896yaj,
    author = {Hogan, C. J.},
    title = {Gravitational radiation from cosmological phase transitions},
    journal = {Monthly Notices of the Royal Astronomical Society},
    volume = {218},
    number = {4},
    pages = {629-636},
    year = {1986},
    month = {02},
    issn = {0035-8711},
    doi = {10.1093/mnras/218.4.629},
    url = {https://doi.org/10.1093/mnras/218.4.629},
    eprint = {https://academic.oup.com/mnras/article-pdf/218/4/629/3299141/mnras218-0629.pdf},
}

@article{LISA:2017pwj,
    author = "Amaro-Seoane, Pau and others",
    collaboration = "LISA",
    title = "{Laser Interferometer Space Antenna}",
    eprint = "1702.00786",
    archivePrefix = "arXiv",
    primaryClass = "astro-ph.IM",
    month = "2",
    year = "2017"
}

@article{Hu:2017mde,
    author = "Hu, Wen-Rui and Wu, Yue-Liang",
    title = "{The Taiji Program in Space for gravitational wave physics and the nature of gravity}",
    doi = "10.1093/nsr/nwx116",
    journal = "Natl. Sci. Rev.",
    volume = "4",
    number = "5",
    pages = "685--686",
    year = "2017"
}

@article{TianQin:2015yph,
    author = "Luo, Jun and others",
    collaboration = "TianQin",
    title = "{TianQin: a space-borne gravitational wave detector}",
    eprint = "1512.02076",
    archivePrefix = "arXiv",
    primaryClass = "astro-ph.IM",
    doi = "10.1088/0264-9381/33/3/035010",
    journal = "Class. Quant. Grav.",
    volume = "33",
    number = "3",
    pages = "035010",
    year = "2016"
}

@article{Musha:2017usi,
    author = "Musha, Mitsuru",
    editor = "Cugny, Bruno and Karafolas, Nikos and Sodnik, Zoran",
    collaboration = "DECIGO Working group",
    title = "{Space gravitational wave detector DECIGO/pre-DECIGO}",
    doi = "10.1117/12.2296050",
    journal = "Proc. SPIE Int. Soc. Opt. Eng.",
    volume = "10562",
    pages = "105623T",
    year = "2017"
}

@article{Corbin:2005ny,
    author = "Corbin, Vincent and Cornish, Neil J.",
    title = "{Detecting the cosmic gravitational wave background with the big bang observer}",
    eprint = "gr-qc/0512039",
    archivePrefix = "arXiv",
    doi = "10.1088/0264-9381/23/7/014",
    journal = "Class. Quant. Grav.",
    volume = "23",
    pages = "2435--2446",
    year = "2006"
}

@article{Steinhardt:1981ct,
    author = "Steinhardt, Paul Joseph",
    title = "{Relativistic Detonation Waves and Bubble Growth in False Vacuum Decay}",
    reportNumber = "UPR-0181T",
    doi = "10.1103/PhysRevD.25.2074",
    journal = "Phys. Rev. D",
    volume = "25",
    pages = "2074",
    year = "1982"
}

@article{Kosowsky:1992rz,
    author = "Kosowsky, Arthur and Turner, Michael S. and Watkins, Richard",
    title = "{Gravitational Waves from First Order Cosmological Phase Transitions}",
    reportNumber = "FERMILAB-PUB-91-333-A-REV, FERMILAB-PUB-91-333-A",
    doi = "10.1103/PhysRevLett.69.2026",
    journal = "Phys. Rev. Lett.",
    volume = "69",
    pages = "2026--2029",
    year = "1992"
}

@article{Megevand:2012rt,
    author = "Megevand, Ariel and Sanchez, Alejandro D.",
    title = "{Analytic approach to the motion of cosmological phase transition fronts}",
    eprint = "1206.2339",
    archivePrefix = "arXiv",
    primaryClass = "astro-ph.CO",
    doi = "10.1016/j.nuclphysb.2012.08.001",
    journal = "Nucl. Phys. B",
    volume = "865",
    pages = "217--237",
    year = "2012"
}

@article{Espinosa:2010hh,
    author = "Espinosa, Jose R. and Konstandin, Thomas and No, Jose M. and Servant, Geraldine",
    title = "{Energy Budget of Cosmological First-order Phase Transitions}",
    eprint = "1004.4187",
    archivePrefix = "arXiv",
    primaryClass = "hep-ph",
    reportNumber = "CERN-PH-TH-2010-027",
    doi = "10.1088/1475-7516/2010/06/028",
    journal = "JCAP",
    volume = "06",
    pages = "028",
    year = "2010"
}

@article{Megevand:2009ut,
    author = "Megevand, Ariel and Sanchez, Alejandro D.",
    title = "{Detonations and deflagrations in cosmological phase transitions}",
    eprint = "0904.1753",
    archivePrefix = "arXiv",
    primaryClass = "hep-ph",
    doi = "10.1016/j.nuclphysb.2009.05.007",
    journal = "Nucl. Phys. B",
    volume = "820",
    pages = "47--74",
    year = "2009"
}

@article{Leitao:2010yw,
    author = "Leitao, Leonardo and Megevand, Ariel",
    title = "{Spherical and non-spherical bubbles in cosmological phase transitions}",
    eprint = "1010.2134",
    archivePrefix = "arXiv",
    primaryClass = "astro-ph.CO",
    doi = "10.1016/j.nuclphysb.2010.11.012",
    journal = "Nucl. Phys. B",
    volume = "844",
    pages = "450--470",
    year = "2011"
}

@article{Leitao:2015fmj,
    author = "Leitao, Leonardo and Megevand, Ariel",
    title = "{Gravitational waves from a very strong electroweak phase transition}",
    eprint = "1512.08962",
    archivePrefix = "arXiv",
    primaryClass = "astro-ph.CO",
    doi = "10.1088/1475-7516/2016/05/037",
    journal = "JCAP",
    volume = "05",
    pages = "037",
    year = "2016"
}

@article{Leitao:2015ola,
    author = "Leitao, Leonardo and Megevand, Ariel",
    title = "{Hydrodynamics of ultra-relativistic bubble walls}",
    eprint = "1510.07747",
    archivePrefix = "arXiv",
    primaryClass = "astro-ph.CO",
    doi = "10.1016/j.nuclphysb.2016.02.009",
    journal = "Nucl. Phys. B",
    volume = "905",
    pages = "45--72",
    year = "2016"
}

@article{Konstandin:2010dm,
    author = "Konstandin, Thomas and No, Jose M.",
    title = "{Hydrodynamic obstruction to bubble expansion}",
    eprint = "1011.3735",
    archivePrefix = "arXiv",
    primaryClass = "hep-ph",
    doi = "10.1088/1475-7516/2011/02/008",
    journal = "JCAP",
    volume = "02",
    pages = "008",
    year = "2011"
}

@article{Giese:2020rtr,
    author = "Giese, Felix and Konstandin, Thomas and van de Vis, Jorinde",
    title = "{Model-independent energy budget of cosmological first-order phase transitions{\textemdash}A sound argument to go beyond the bag model}",
    eprint = "2004.06995",
    archivePrefix = "arXiv",
    primaryClass = "astro-ph.CO",
    reportNumber = "DESY-20-064",
    doi = "10.1088/1475-7516/2020/07/057",
    journal = "JCAP",
    volume = "07",
    number = "07",
    pages = "057",
    year = "2020"
}

@article{Giese:2020znk,
    author = "Giese, Felix and Konstandin, Thomas and Schmitz, Kai and van de Vis, Jorinde",
    title = "{Model-independent energy budget for LISA}",
    eprint = "2010.09744",
    archivePrefix = "arXiv",
    primaryClass = "astro-ph.CO",
    reportNumber = "DESY-20-173, DESY 20-173, CERN-TH-2020-170",
    doi = "10.1088/1475-7516/2021/01/072",
    journal = "JCAP",
    volume = "01",
    pages = "072",
    year = "2021"
}

@article{Wang:2020nzm,
    author = "Wang, Xiao and Huang, Fa Peng and Zhang, Xinmin",
    title = "{Energy budget and the gravitational wave spectra beyond the bag model}",
    eprint = "2010.13770",
    archivePrefix = "arXiv",
    primaryClass = "astro-ph.CO",
    doi = "10.1103/PhysRevD.103.103520",
    journal = "Phys. Rev. D",
    volume = "103",
    number = "10",
    pages = "103520",
    year = "2021"
}

@article{Tian:2024ysd,
    author = "Tian, Chi and Wang, Xiao and Bal{\'a}zs, Csaba",
    title = "{Gravitational waves from cosmological first-order phase transitions with precise hydrodynamics}",
    eprint = "2409.14505",
    archivePrefix = "arXiv",
    primaryClass = "hep-ph",
    doi = "10.1140/epjc/s10052-025-14826-2",
    journal = "Eur. Phys. J. C",
    volume = "85",
    number = "10",
    pages = "1091",
    year = "2025"
}

@article{Yamada:2025cfr,
    author = "Yamada, Masaki",
    title = "{Analytic derivation of the GW spectrum from bubble collisions in an FLRW universe}",
    eprint = "2509.16073",
    archivePrefix = "arXiv",
    primaryClass = "astro-ph.CO",
    reportNumber = "TU-1276",
    doi = "10.1103/p5h3-zrsx",
    journal = "Phys. Rev. D",
    volume = "113",
    number = "2",
    pages = "023517",
    year = "2026"
}

@article{Yamada:2025hfs,
    author = "Yamada, Masaki",
    title = "{Maximal GW amplitude from bubble collisions in supercooled phase transitions}",
    eprint = "2509.13402",
    archivePrefix = "arXiv",
    primaryClass = "gr-qc",
    reportNumber = "TU-1275",
    doi = "10.1016/j.physletb.2026.140157",
    journal = "Phys. Lett. B",
    volume = "873",
    pages = "140157",
    year = "2026"
}

\end{document}